\documentclass[preprint,5p,times,twocolumn]{elsarticle}

\usepackage{amssymb}
\usepackage{verbatim}
\usepackage[linesnumbered,ruled,vlined]{algorithm2e}
\usepackage{graphicx}
\usepackage{multirow}
\usepackage{amsfonts}
\usepackage[caption=false]{subfig}
\usepackage{textcomp}
\usepackage{stfloats}
\usepackage{url}
\usepackage{hyperref}
\usepackage{balance}
\usepackage{amsmath}
\usepackage{pifont}
\usepackage{booktabs} % 用于三线表
\usepackage{tikz}
\usepackage{xcolor}
\usepackage{enumitem}
\definecolor{darkgreen}{rgb}{0.0, 0.39, 0.0} % 定义深绿色

\newcommand{\fullcircle}{%
  \begin{tikzpicture}
    \fill[black] (0.06,0) arc[start angle=0,end angle=180,radius=0.5ex] -- (0,0) -- cycle;
    \fill[black] (-0.06,0) arc[start angle=180,end angle=360,radius=0.5ex] -- (0,0) -- cycle;
    \draw (0,0) circle[radius=0.5ex];
  \end{tikzpicture}%
  \hspace{0.3ex} % 调整间距
}
\newcommand{\halfcircle}{%
  \begin{tikzpicture}
    \fill[black] (0.06,0) arc[start angle=0,end angle=180,radius=0.5ex] -- (0,0) -- cycle;
    \fill[white] (-0.06,0) arc[start angle=180,end angle=360,radius=0.5ex] -- (0,0) -- cycle;
    \draw (0,0) circle[radius=0.5ex];
  \end{tikzpicture}%
  \hspace{0.3ex} % 调整间距
}
\journal{Computer $\&$ Security}

\begin{document}

\begin{frontmatter}



\title{\textsc{ProHunter}: A Comprehensive APT Hunting System Based on Whole-System Provenance}


\author[1]{Xuebo Qiu}

\author[2]{Mingqi Lv\corref{cor1}}
\ead{mingqilv@zjut.edu.cn}

\author[1]{Tiantian Zhu}
\author[1]{Yimei Zhang}
\author[2]{Tieming Chen}

%\address[1]{Zhejiang University of Technology, China}
\address[1]{College of Computer Science and Technology, Zhejiang University of Technology, Hangzhou, 310023, China}
\address[2]{College of Geoinformatics, Zhejiang University of Technology, Hangzhou, 310023, China}
\cortext[cor1]{Corresponding author}

% \affiliation{organization={},%Department and Organization

\begin{abstract}
  Advanced Persistent Threats (APTs) remain difficult to detect due to their stealthy nature and long-term persistence. To tackle this challenge, provenance-based threat hunting has gained traction as a proactive defense mechanism. This technique models audit logs as a whole-system provenance graph and searches for subgraphs that match APT patterns recorded in Cyber Threat Intelligence (CTI) reports.
  However, several limitations persist: 1) significant memory and time overhead due to the extremely large provenance graphs; 2) imprecise segmentation of APT activities from provenance graphs due to their intricate entanglement with benign operations; and 3) poor alignment of attack representations between CTI-derived query graphs and provenance graphs due to their substantial semantic gaps.
  
  To address these limitations, this paper presents \textsc{ProHunter}, an efficient and accurate provenance-based APT hunting system with a platform-independent design.
  To minimize system overhead, \textsc{ProHunter} creates a compact data structure that efficiently stores long-term provenance graphs using semantic abstraction and bit-level hierarchical encoding strategies.
  To segment APT behaviors, a heuristic-driven threat graph sampling algorithm is designed, which can extract precise attack patterns from provenance graphs. 
  Furthermore, to bridge the semantic gaps between CTI-derived graphs and provenance graphs, \textsc{ProHunter} proposes adaptive graph representation and feature enhancement methods, enabling the extraction of consistent attack semantics at both localized and globalized levels.
  Extensive evaluations on real-world APT campaigns from DARPA TC E3, E5 and OpTC datasets demonstrate that \textsc{ProHunter} outperforms state-of-the-art threat hunting systems in terms of efficiency and accuracy. Our code is available at \url{https://github.com/xueboQiu/ProHunter}.

\end{abstract}


% \begin{highlights}
% \end{highlights}

\begin{keyword} Advanced Persistent Threats \sep
  Data Provenance \sep
  Threat Hunting

\end{keyword}

\end{frontmatter}


\section{Introduction} \label{introduction}
{\color{black}
Advanced Persistent Threats (APTs) remain one of the most critical challenges in modern cybersecurity. Their staged execution, ``low-and-slow'' patterns, and ability to blend into normal system activities allow adversaries to keep undetected for extended periods. Consequently, defenders increasingly rely on \textit{threat hunting}, a proactive and intelligence-driven technique where analysts search for traces of known or suspected attack behaviors within hosts to shorten APT dwell time \cite{crowdstrike}.

Provenance-based threat hunting, which integrates whole-system provenance capture with Cyber Threat Intelligence (CTI) knowledge, has emerged as a promising direction \cite{megrapt,deephunter,poirot,ma2025actminer,provgsearcher}. 
By transforming fine-grained audit logs into directed provenance graphs that encode causal relationships among processes, files, and network flows, it enables analysts to reconstruct attack paths and identify malicious intent with high fidelity. 
Building on this structured execution history, the system applies CTI knowledge, such as Indicators of Compromise (IoC), to search for matching patterns and expose intrusions missed by traditional detection. 
In principle, the combination of system provenance and CTI promises actionable threat hunting. However, realizing this paradigm remains far from operationally effective due to three core challenges (C1-C3):

\indent\textbf{\textit{C1: Efficient Memory Utilization.}}
Real-world systems generate daily provenance graphs with millions of nodes and edges, many carrying memory-intensive attributes. Production environments, however, often allocate available memory resources for security analytics (e.g., $<$ 100 MB/host  \cite{arewethere}). 
The tension between massive graph analysis and tight resource budgets prevents these systems from continuously running at scale, and creates a bottleneck for efficient storage and management.

\textbf{\textit{C2: Attack Activity Sampling.}} 
Malicious events are rare and tightly interwoven with benign activities. Therefore, isolating a threat graph (i.e., a subgraph encompassing attack-related activities) is crucial for focused threat analysis. However, adversaries actively mimic normal behaviors \cite{poirot}, and the phenomenon of dependency explosion, where persistent entities like \texttt{explorer.exe} accumulate extensive dependencies over time, further exacerbates the entanglement between malicious and benign interactions. 
This complexity impedes the precise threat graph sampling and effective downstream threat analysis.

\textbf{\textit{C3: Bridging Semantic Gaps.}} 
CTI reports describe attack behaviors in high-level domain terminology (e.g., ``conduct ARP scan''), whereas audit logs capture low-level system events (e.g.,``\texttt{cmd.exe} execution''). This mismatch produces inconsistent attack  interpretations between CTI and provenance graphs, leading to poor CTI utilization during threat hunting \cite{ma2025actminer}.

\begin{figure*}[t!]
  \centering
  \color{black}
  \includegraphics[scale=1.25]{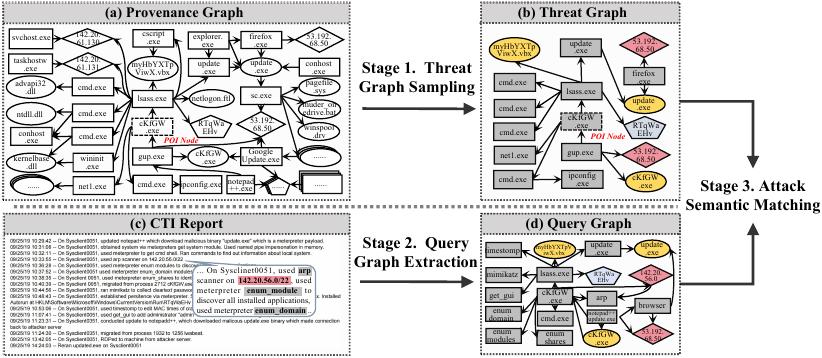}
  \caption{Overview of the three-stage threat hunting pipeline. The attack campaign shown is ``Malicious Escalation'' from Day 3 of the OpTC dataset. Stage 1 samples a threat graph (b) from a provenance graph (a). Stage 2 extracts a query graph (d) from a CTI report (c). Stage 3 performs attack semantic matching between query and threat graphs to identify threats. Node shapes represent entity types: rectangles (processes), ellipses (files), diamonds (netflows), and pentagons (registries). Colored nodes in (b) and (d) represent matched attack entities, while highlighted entities in (c) corresponds to extracted nodes in (d).}
  \label{challenge}
\end{figure*}

Current provenance-based threat hunting systems tackle these challenges in isolation, which results in fragile solutions \cite{provgsearcher,megrapt,deephunter,ghunter,poirot,threatraptor,ma2025actminer}. In practice, they either fail under tight resource limits \cite{poirot,ma2025actminer,deephunter}, miss attacks when explicit IoCs are unavailable \cite{megrapt,threatraptor}, or produce unstable matches due to unresolved semantic gaps \cite{ghunter,provgsearcher}. These weaknesses prevent current systems from delivering reliable threat hunting. This core gap motivates our work.
To ground this problem, the ``Malicious Escalation'' campaign in the DARPA OpTC dataset \cite{darpatcoptc} is used as a case to illustrate the standard three-stage threat hunting pipeline employed by prior systems, and reveals the specific points where they break down. The working pipeline is shown in Figure~\ref{challenge}.
\begin{itemize}[leftmargin=15pt]
  \item \textbf{Stage 1: Threat Graph Sampling.} Identify Point of Interest (POI) nodes (e.g., the process \texttt{cKfGW.exe} in Figure~\ref{challenge}a) via IoC matching \cite{megrapt} or anomaly detection \cite{deephunter}, and use them as anchors to sample a threat graph (Figure \ref{challenge}b).
  \item \textbf{Stage 2: Query Graph Extraction.} Translate CTI reports (Figure~\ref{challenge}c) into structured query graphs (Figure~\ref{challenge}d) that encode key malicious actions and dependencies in a format aligned with provenance graphs.
  \item \textbf{Stage 3: Attack Semantic Matching.} Match the sampled threat graph against query graphs to infer attack presence.
\end{itemize}

Although well-established, this pipeline exposes four practical limitations of existing systems.
(1) \textbf{Efficiency}. 
Most prior solutions \cite{deephunter,poirot} require repeatedly preloading of provenance graphs through cumbersome graph-modeling frameworks (e.g., NetworkX \cite{networkx}), which introduces high latency and cost during both Stages 1 and 3.
(2) \textbf{Adaptability}.
Attackers increasingly exploit stealthy attack vectors (e.g., fileless attacks \cite{filelesssurvey}) to evade detection. Systems \cite{deephunter,megrapt} that depend on explicit POIs (e.g., IoCs) for threat graph sampling in Stage 1 often fail, as the required anchor signals may not exist. 
(3) \textbf{Accuracy}. 
Semantic gaps plague current semantic matching strategies \cite{provgsearcher,ghunter}. For example, the execution of \texttt{enum modules} in the query graph (Figure \ref{challenge}d) may appear as a \texttt{cmd.exe} process in the threat graph (Figure \ref{challenge}b), while behaviors like \texttt{arp} scan lack direct counterparts. Existing systems reconcile these asymmetric semantics by lifting node semantics to coarse-grained node types in Stage 2 and matching node-type interactions between the query graph and the threat graph in Stage 3, but this coarse abstraction ultimately produces a high false positive rate.
(4) \textbf{Scalability}. 
APT campaigns may endure for up to 90 days \cite{sans}, generating enormous volumes of audit logs. 
Existing practices often require gigabytes of memory to load even a small subset of this data \cite{poirot,deephunter}, rendering long-term analysis unscalable for Stages 1 and 3.

To address these issues, this paper presents \textsc{ProHunter}, a comprehensive threat hunting system that fulfills these practical requirements by tackling the aforementioned challenges.
For challenge C1, \textsc{ProHunter} designs a compact in-memory provenance graph structure that applies semantic abstraction and hierarchical encoding, enabling bit-level compression of system nodes and edges according to their characteristics for highly efficient memory utilization.
For challenge C2, the system devises a heuristic Breadth-First Search (BFS) algorithm that integrates propagation-aware sampling rules to track suspicious information flows without relying on explicit POI signals, thereby capturing contextually anomalous interactions, particularly those obscured by explosive dependencies.
For challenge C3, \textsc{ProHunter} proposes a customized Graph Isomorphism Network (GIN) \cite{gin}, integrated with inter-graph message passing and feature enhancement strategies, to jointly capture both localized and globalized consistencies in attack semantics between query graphs and threat graphs.
}
%We implement \textsc{ProHunter} and open-source for further research. 

Comprehensive experiments were conducted to evaluate \textsc{ProHunter} across memory utilization, threat graph sampling precision, and threat hunting performance using the DARPA TC E3 \cite{darpatce3}, E5 \cite{darpatce5}, and OpTC \cite{darpatcoptc} datasets. To the best of our knowledge, these datasets constitute the most comprehensive benchmark suite for threat hunting evaluation, encompassing a total of 28 attack scenarios. The experimental results demonstrate the exceptional efficiency of \textsc{ProHunter}, requiring only 3–25 MB/day to store provenance graphs. Its sampled threat graphs exhibit over 70\% attack semantic coverage while introducing merely 2–3 irrelevant nodes on average, significantly outperforming prior methods \cite{watson,deephunter,megrapt}. {\color{black} In threat hunting, \textsc{ProHunter} sets a new benchmark, attaining 100\% recall with an average false alarm rate below 1\%. Notably, it operates effectively even without POI signals while maintaining a clear separation between threat and benign behaviors.}
%nd comparing it with MEGR-APT and the subgraph sampling algorithms in Watson, we validate that our sampling approach excels in semantic coverage of suspicious behaviors as well as irrelevant entity coverage. Finally, by comparing the accuracy of MEGR-APT,ProvG-Searcher hunting in different scenarios in DARPA TC \cite{darpatce3,DARPAtcoptc}, it is found that \textsc{ProHunter} possesses higher accuracy as well as robustness, as well as lower false alarm rate when dealing with benign suspicious subgraphs. The higher robustness performance of our method is verified. Meanwhile, through performance comparison, we find that our method can determine whether a suspicious picking subgraph is a known threat attack graph in less than 2s after getting POI hints, which is improved by x180 compared with MEGR-APT as well as Poirot. which verifies the timeliness of our method.
Our contributions are summarized as follows:
\begin{itemize}
  \item {\color{black} The paper presents \textsc{ProHunter}, a comprehensive threat hunting system that synergistically integrates provenance graph compaction, threat graph sampling, and attack representation and matching to advance APT hunting capabilities.}
  \item The paper designs a memory-efficient provenance graph storage structure based on semantic abstraction and hierarchical encoding, offering critical insights into abstract node types and suspicious information flows.
  \item {\color{black} The paper develops a heuristic subgraph sampling algorithm that accommodates diverse POI assumptions (e.g., arbitrary system nodes), extending the practical applicability of threat hunting.}
  %\item[-] We develop a heuristic sampling algorithm based on abstracted node types and suspicious information flows to precisely identify threat interactions. particularly those buried in explosive dependencies.}
  \item The paper introduces an adaptive graph representation and enhanced feature extraction strategies to bridge the semantic gaps between provenance graphs and CTI-derived query graphs.
  \item The paper evaluates \textsc{ProHunter} on multiple DARPA datasets, demonstrating its superiority over leading threat hunting systems in generality, efficiency and accuracy.
\end{itemize}

\section{Background}
\subsection{Definitions} \label{query generate}
\subsubsection{Provenance Graph} A provenance graph represents audit logs as a multi-directed and attributed graph that captures the causal dependencies of system entities. Formally, it is defined as $\mathcal{G}=(\mathcal{V}, \mathcal{E})$, where $\mathcal{V}$ represents system entities, and $\mathcal{E}$ consists of edges representing interactions between system entities. Each event is denoted as $<sbj, op, obj, ts>$, where $sbj$ and $obj$ are the subject and object of the event, $op$ indicates the edge type (e.g., read), and $ts$ is the event timestamp. Table \ref{nodes edges} lists our monitored node and edge types.

\subsubsection{Query Graph}
A query graph encodes prior attack knowledge recorded in CTI reports (e.g., after-action reports from DARPA \cite{darpatce3,darpatce5,darpatcoptc}), as illustrated in Figure \ref{challenge}d. 
Since CTI reports are often unstructured and verbose, directly translating them into usable graphs poses significant challenges. To this end, \textsc{ProHunter} adopts a semi-automated construction pipeline. First, raw query graphs are automatically extracted from CTIs using AttacKG \cite{attackg}, and then refined through manual correction and complementation.

\subsubsection{Points of Interest (POIs)}
In threat hunting, POIs serve as entry points within provenance graphs for locating threat graphs. Prior work  typically selects POIs flagged by IoC matching \cite{megrapt} or anomaly detection alerts \cite{deephunter} to narrow the search space. In contrast, \textsc{ProHunter} generalizes the POI definition to encompass arbitrary system nodes, providing a more robust approach that eliminates reliance on pre-existing indicators.

\subsubsection{Problem and Goal} 
\textsc{ProHunter} treats threat hunting as a graph matching task \cite{megrapt,deephunter}. Formally, given a query graph $\mathcal{G}_q$ = ($\mathcal{V}_q$, $\mathcal{E}_q$) and a provenance graph $\mathcal{G}_p$, \textsc{ProHunter} aims to search for a threat graph $\mathcal{G}_{t}$  = ($\mathcal{V}_t$, $\mathcal{E}_t$) $\subseteq$ $\mathcal{G}_{p}$ that is isomorphic to $\mathcal{G}_q$, i.e., $\exists$ a bijection \( f : \mathcal{V}_t \rightarrow \mathcal{V}_q \) such that for all \( u, v \in \mathcal{V}_t \): \( (u, v) \in \mathcal{E}_t \iff (f(u), f(v)) \in \mathcal{E}_q\).
There are two primary techniques for graph matching: \textit{exact matching} and \textit{approximate matching}.
\textsc{ProHunter} adopts the approximate matching technique for the following reasons: 1) exact matching is NP-complete and incurs significant computational overhead, making it impractical for large-scale and real-time threat hunting scenarios \cite{egm}; and 2) approximate matching is better suited for scenarios where graphs are structurally similar but not perfectly isomorphic, offering greater resilience to semantic gaps and adaptability against APT attacks.

\begin{table}[]
  \centering
  \caption{System entities and edges in the provenance graph.}
  \label{nodes edges}
  \scriptsize
  \begin{tabular}{|l|l|}
  \hline
  \textbf{Node Pair} & \textbf{Edge Type}                                      
  \\ \hline  \multicolumn{1}{l}{} \vspace{-7pt} \\\hline
  Process$\rightarrow$Process   & Fork,Exec,Modify,Open         \\ \hline
  Process$\rightarrow$File      & \begin{tabular}[l]{@{}l@{}}Create,Read,Write,Rename,Link,\\ Unlink,Modify,Delete,Load\end{tabular} \\ \hline
  Process$\rightarrow$NetFlow   & Connect,Start,Send,Recv,Message                         \\ \hline
  \end{tabular}
  \end{table}
\begin{table}[]
  \caption{Comparison of threat hunting systems. Half-filled circle indicates partial fulfillment of the requirement, while fully-filled circle denotes complete fulfillment.}
  \centering
  \label{metric compare}
  \scriptsize
  \setlength{\tabcolsep}{3.7pt} % 减少列间距
\begin{tabular}{|l|c|c|c|c|}
  \hline
  \textbf{Scheme}& \textbf{Efficiency} & \textbf{Adaptability} & \textbf{Accuracy} &
  \textbf{Scalability}
  \\ \hline  \multicolumn{1}{c}{} \vspace{-6.5pt} \\\hline
  Poirot \cite{poirot}                      &              & \fullcircle            & \fullcircle            &             \\ \hline
  MEGR-APT \cite{megrapt}                     &              &             &\halfcircle            & \fullcircle            \\ \hline
  ProvG-Searcher \cite{provgsearcher}                       & \fullcircle             & \halfcircle            & \halfcircle            & \halfcircle           \\ \hline
  GHunter  \cite{ghunter}                     & \fullcircle             & \halfcircle            &           &             \\ \hline
  DeepHunter \cite{deephunter}                   &             & \halfcircle & \halfcircle                          &             \\ \hline
  \textbf{\textsc{ProHunter}} &\fullcircle             & \fullcircle             & \fullcircle             & \fullcircle            \\ \hline
\end{tabular} 
  \end{table}

\subsection{Previous Research Limitation}
This section compares \textsc{ProHunter} with existing threat hunting solutions \cite{megrapt,deephunter,ghunter,provgsearcher,poirot} across the practical dimensions outlined in Section \ref{introduction}, with the results summarized in Table~\ref{metric compare}.

\subsubsection{Efficiency}
{\color{black}
MEGR-APT \cite{megrapt} requires one hour per day to parse audit logs and store the provenance graph in an RDF-based database, which results in a time lag. Additionally, it takes minutes to sample threat graphs using SPARQL query statements and perform threat hunting. Poirot \cite{poirot} and DeepHunter \cite{deephunter} rely on preloading provenance graphs using heavy data structures (e.g., NetworkX \cite{networkx}), which are both memory-intensive (GB/day) and time-consuming. 
These solutions fail to satisfy the efficiency requirement. 
Conversely, \textsc{ProHunter} introduces a memory-efficient data structure to represent provenance graphs, enabling real-time threat graph sampling and threat hunting within seconds. 

\subsubsection{Adaptability} 
ProvG-Searcher \cite{provgsearcher} and GHunter \cite{ghunter} employ the neural subgraph prediction method based on order embedding techniques \cite{neuro} to conduct threat hunting. However, as the provenance graph expands, the increasing number of subgraph patterns may generate false matches to query graphs, potentially causing alert fatigue \cite{nodoze}.
DeepHunter and MEGR-APT sample threat graphs using a set of IoCs (e.g., \texttt{cKfGW.exe} in Figure \ref{challenge}b) as POI signals. However, as defined in Pyramid of Pain \cite{pyramidofpain}, attackers can easily manipulate IoCs, thereby undermining these solutions. 
\textsc{ProHunter} extracts threat graphs by focusing on information flows and abstracted node semantics, eliminating the reliance on explicit signatures and enabling flexible adaptation to diverse POIs.

% 系统架构图
\begin{figure*}[]
  \centering
  \includegraphics[scale=0.72]{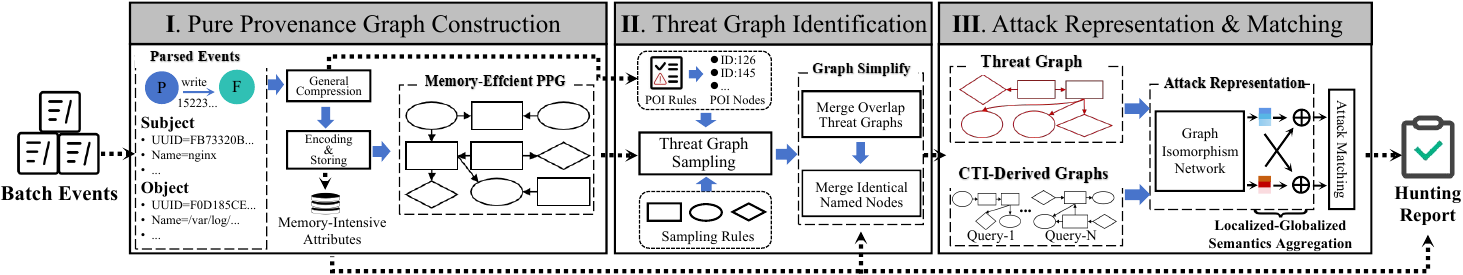}
  \caption{System architecture of \textsc{ProHunter}.}
  \label{framework}
\end{figure*}

\subsubsection{Accuracy}
The semantic gaps illustrated in Figure \ref{challenge} and potential attack perturbations \cite{mimicry} can compromise the subgraph relationships, resulting in subgraph entailment mispredictions by ProvG-Searcher and GHunter. 
Moreover, MEGR-APT and DeepHunter struggle to distill deeper attack semantics due to their simplistic feature extraction and representation methods, resulting in suboptimal accuracy.
In contrast, \textsc{ProHunter} generates enriched attack representations through an enhanced feature extraction method based on semantic abstractions and adaptive graph representation strategies, effectively capturing both localized and globalized attack semantics shared between threat and query graphs.
}

\subsubsection{Scalability} 
Poirot and DeepHunter require loading the whole provenance graph into memory to optimize graph matching efficiency, leading to significant overhead. Similarly, ProvG-Searcher and GHunter need to frequently predict subgraph relationships between query graphs and evolving provenance graphs, incurring substantial computational costs. \textsc{ProHunter} features a memory-efficient PPG structure with low overhead, and a precise threat graph sampling algorithm that narrows the hunting scope, thereby enhancing scalability.

\subsection{Threat Model}
%Since \textsc{ProHunter} identifies lurking intruders based on audit logs, 
This paper assumes the integrity of the underlying system, treating the operating system, auditing framework, and system logs as part of the Trusted Computing Base (TCB), in line with  prior work \cite{holmes,aptkgl,threatrace,magic}. The detection of side-channel attacks, covert channel attacks and kernel exploits fall outside the scope of this research. Moreover, \textsc{ProHunter} presumes the reliability of query graphs used for threat hunting. Therefore, the extraction and validation of these graphs from CTI reports are considered out-of-scope, as these topics constitute a distinct research area \cite{attackg,extractor}. 
Regarding the adversary, we model a sophisticated attacker capable of obfuscating their actions, but we posit that their core tactics remain fundamentally consistent with the patterns documented in CTI reports, aligning with prior studies~\cite{provgsearcher,megrapt}.

\section{System Design}
\subsection{Overview}
\textsc{ProHunter} achieves efficient and accurate APT hunting based on approximate graph matching mechanism, which comprises the following components: 1) Pure Provenance Graph (PPG) construction; 2) threat graph sampling; and 3) attack representation and matching.

The pipeline is shown in Figure \ref{framework}. 
First, the PPG construction component applies existing general compression strategies to deduplicate the batch audit logs and then 
%removing consecutive duplicate events \cite{gsss} and identical network events within fixed time windows \cite{versioning}, detailed in Appendix \ref{generic_compression}. After deduplication, it 
performs semantic abstraction and hierarchical encoding to dynamically construct the PPG.
%Those memory-intensive attributes are stored into the database for subsequent attack graph reconstruction. 
Second, the threat graph sampling algorithm uses a set of heuristic rules to sample threat graphs from the PPG centered on POI nodes.
Finally, the attack representation and matching component initializes graph features, extracts consistent localized and globalized attack semantics between threat and query graphs, and performs approximate graph matching based on enhanced attack representations. 
%Following a successful threat hunting, \textsc{ProHunter} retrieves the full attributes of the threat graph from databases for visualization.
%These identified graphs are then delivered to the PPG graphical module to visualise, thereby assisting security analysts in conducting attack investigations. 
%Each component is described in detail as follows.

\subsection{Pure Provenance Graph Construction}\label{sec:ppg}

Provenance graphs that model complete system interactions facilitate threat analysis.
Previous methods store provenance graphs in databases and analyze through database engines \cite{megrapt}, which introduces significant latency and affects efficiency. Alternative methods \cite{poirot, deephunter} leverage graphical libraries to process provenance graphs, resulting in considerable memory usage and reduced scalability (Challenge C1). 
Aiming at this challenge, Sleuth \cite{sleuth} introduces a variable-length encoding strategy to compact storage of provenance graphs. However, several shortcomings remain: 
1) it only compacts edges while neglecting node compaction, where the amount of edges and nodes in the deduplicated provenance graph is within \textit{the same order of magnitude} (e.g., 2x in the OpTC dataset, detailed results are provided in \ref{generic_compression}), underscoring the importance of node compaction; and 2) its encoding strategy partitions continuous memory space into variable-length segments, which complicates the decoding process and disrupts memory alignment, ultimately diminishing memory access efficiency \cite{mem_align}.
PPG addresses these issues using two dependency-preserving compaction strategies: semantic abstraction and hierarchical encoding. 

\subsubsection{Storage Architecture} 
\begin{figure}[t]
  \centering
  \includegraphics[width=3in]{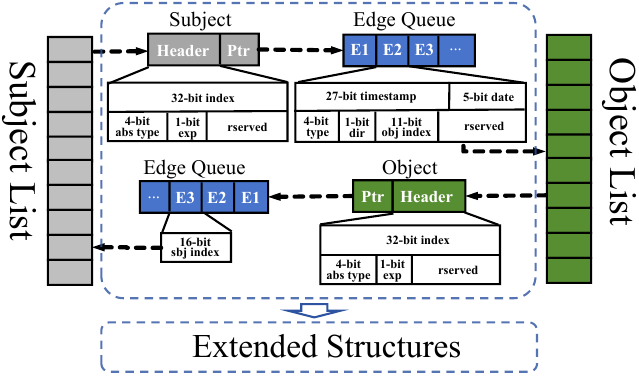}
  \caption{\centering{The storage architecture of PPG.}}
  \label{ppg}
\end{figure}

%The structures of PPG are also encoded at the \textit{bit level} and leverage 32-bit pointers instead of 64-bit ones to minimize memory waste.
%We use two fixed-size structures to store nodes with their dependencies with different characteristics to ensure efficient memory access efficiency.
%The employment of disparate levels of compression methodologies enables the effective utilization of memory space, while ensuring the preservation of dependency relationships. 
%The provenance graph constructed by PPG contains the nodes and edges listed in Table  \ref{nodes edges}, which are more audit-worthy. 
%which provide meaningful semantics.
The architecture of PPG, illustrated in Figure \ref{ppg}, consists of three types of elements: Subject, Object, and Edge, all encoded at the bit level.
Each subject and object node is stored in a separate list, where each node includes a header storing attributes and a pointer to a queue of connected edges. 
%The subject node is stored in a subject list, with each node allocated a 32-bit pointer to the queue of connected edges. Similarly, each object is stored in an object list, with a 32-bit pointer to its connected subject queue, containing subject identifiers for backward tracing. 
The subject's edge queue stores both edge attributes and identifiers of connected objects, whereas the object's edge queue only contains identifiers of connected subjects, aiming to minimize the redundant edge storage.
This decoupled design allows the use of customized variable-sized blocks to store different elements, optimizing memory utilization and supporting bidirectional graph traversal. The encoding details are as follows. 

% and 3) it overlooks the semantics of inter-entity interactions, leading to redundant encoding of events. 

%cascading storage strategy based on variable-length encoding. 
%Our observations indicate that the majority of nodes in the origin graph exhibit low degrees of connected, whereas a small number of explosion-dependent nodes demonstrate high degrees of dense connected. Based on this insight, we have designed the hierarchical coding strategy.
\subsubsection{Semantic Abstraction} \label{sec:semantic align}
Node attributes consume substantial memory. In the DARPA TC dataset, each node requires an average of 56 bytes to store attributes (i.e., identifier, name, and type), imposing a huge burden on the memory. To mitigate this issue, we propose a semantic abstraction strategy that replaces the original node type and name with a 4-bit abstract type (see Table \ref{absnodes}) named `abs type' and substitutes the 128-bit identifier with a 32-bit integer index named `index' (see Figure \ref{ppg}). The rationale behind this design is twofold.
\begin{enumerate}
  \item \textbf{Semantic Alignment:} 
  Using low-level node names as node features can lead to inconsistent attack representations for threat hunting, as biased node descriptions across CTIs and audit logs \cite{deephunter}.
  Using high-level node types, conversely, provides insufficient details for distinguishing similar benign behavior patterns \cite{megrapt}. 
  Therefore, we adopt a compromise by converting the original node names and types in both provenance and query graphs into \textit{intermediate-level} abstractions, which align the intended semantics while preserving differentiation from benign behaviors. For example, both the \texttt{enum modules} process in Figure \ref{challenge}b and the \texttt{cmd} process in Figure \ref{challenge}d can be aligned to the abstract type of `util\_process'.
  \item \textbf{Space Efficiency:} The essential semantics for threat hunting are maintained while extraneous details are removed. Moreover, using integer indices for nodes allows referencing connected nodes by their \textit{relative index distances} rather than explicit pointers, thereby further reducing memory usage. For instance, when the process \texttt{notepad++.exe} spawns \texttt{gup.exe}, an edge is added to the edge queue of \texttt{notepad++.exe} with its `obj index' field  (see Figure \ref{ppg}) set to the index difference between the two processes, enabling \texttt{notepad++.exe}  to track the dependency via the `obj index'.
\end{enumerate}

\begin{table}[t]
  \caption{List of abstract node types.}
  \label{absnodes}
  \centering
  \scriptsize
  \begin{tabular}{|l|l|}
  \hline
  \textbf{Node} & \textbf{Abstract Node Type}              
  \\ \hline  \multicolumn{1}{c}{} \vspace{-7pt} \\\hline
  Process            & 
  \begin{tabular}[l]{@{}l@{}} sys\_process, usr\_process, serv\_process, util\_process, \\web\_process, unknown\_process \end{tabular}\\ \hline
  File               & 
  \begin{tabular}[l]{@{}l@{}}lib\_file, sys\_file, cfg\_file, usr\_file, tmp\_file, unknown\_file  \end{tabular}    \\ \hline
  Netflow            & private\_netflow, public\_netflow                  \\ \hline
  \end{tabular}
  \end{table}

\subsubsection{Hierarchical Encoding}
We observe that the majority of nodes in the provenance graph have scarce dependencies (e.g., over 99\% of nodes in the OpTC dataset have fewer than 10 edges), known as \textit{sparse nodes}. These nodes exhibit strong spatio-temporal localization characteristics, i.e., their dependencies are distributed within short timeframes and are not connected to nodes with distant indices.
%(i.e., small index differences). 
This feature ensures that their connected nodes are in close index proximity, allowing for more efficient storage with fewer bits.
Conversely, a small subset of nodes have dense dependencies, either due to having many neighbors or large dependency distances. These nodes, known as \textit{dependency explosion nodes}, require more bits to store the relative indices of their connected nodes. 

These observations motivate the hierarchical encoding for provenance graphs. 
Specifically, the edge queue size for a sparse node is limited to 16. Each edge in the subject's edge queue is assigned an 11-bit `obj index' to store the relative index of the connected object, allowing a maximum index distance of 1024. Similarly, each edge in the object's edge queue is assigned a 16-bit `sbj index'.
When a sparse node's edge queue exceeds the predefined limit or the distance of any dependency surpasses the bit capacity, its 1-bit `exp' flag is activated, promoting it to a dependency explosion node. 
At this point, all its dependencies are migrated to an extended structure with a larger edge queue and memory allocation (e.g., 27-bit `obj index' and 32-bit `sbj index') to accommodate more and distant dependencies. 
This strategy optimizes memory utilization by allocating optimal sized memory blocks for nodes and edges according to their characteristics.
Fixed miscellaneous bits of edges (e.g. 4-bit 'type') are similar to those in nodes.
%This strategy maintains the majority of nodes and their dependencies in compact structures, while nodes with explosive dependencies are accommodated in larger structures, ensuring efficient memory usage.
Due to space limitations, functional extension of the PPG using reserved bits is described in \ref{misc}.

{\color{black}
\subsubsection{Complexity Analysis for PPG Construction.}
This section analyzes the complexity of the PPG construction.
Let $\mathcal{V}$ and $\mathcal{E}$ denote the system entities and events, PPG processes each event in $\mathcal{O}(1)$ via hash-based indexing and queue appending. When a sparse node is promoted to its extended structure, the time cost is $\mathcal{O}(d)$ for a node with degree $d$. Since each node can be promoted at most once and $\sum_{v \in \mathcal{V}} d_v = 2|\mathcal{E}|$, the total promotion overhead is bounded by $\mathcal{O}(|\mathcal{E}|)$. Therefore, PPG supports linear-time construction with total complexity $\mathcal{O}(|\mathcal{E}|)$. The space complexity is $\mathcal{O}(|\mathcal{V}| + |\mathcal{E}|)$, demonstrating that semantic abstraction and hierarchical encoding enables storage of massive provenance graphs within tight memory budgets.
}

\subsection{Threat Graph Sampling} \label{sec:sampling}
A well-sampled threat graph can significantly enhance threat hunting and security analysis.
Existing methods \cite{megrapt,deephunter}, however, rely on explicit IoCs to sample threat graphs, rendering them vulnerable to evasion techniques \cite{pyramidofpain}. Furthermore, they rigidly restrict the number of sampled nodes, neglecting suspicious dependencies obscured by dependency explosion nodes. This results in attack semantic loss in the sampled threat graph (Challenge C2). 
{\color{black}
To overcome these limitations, we shift from explicit IoCs to broader behavior patterns by correlating information flows and abstracted node semantics within the PPG structure. The behaviors modeled by the PPG encapsulate valuable insights into suspicious flows \cite{holmes, aptshield}. 
For instance, tracking information flows from a file with the abstracted type `sys\_file' to a `public\_netflow' address via an unexpected `usr\_process' process can reveal potential exfiltration of system information. 
Building on this principle, we develop an adaptive BFS algorithm to identify covert malicious patterns by tracing the propagation of suspicious information flows across abstracted nodes, while systematically filtering out benign interactions. This propagation-aware strategy supports threat graph sampling under broader threat hypotheses, such as specifying POIs as arbitrary system entities.
} 
\subsubsection{Sampling Procedure} 
The sampling algorithm takes the constructed PPG, a sampling hop $k$, and POIs as input, with the pseudocode outlined in \ref{app:adaptive bfs}. 
%It receives POI event associated entities as seed nodes due to their higher suspicion, and a predefined traversal depth $k$ as inputs, where 
It first designates the POIs as seed nodes,
%(line 4)
and then conducts a bidirectional recursive traversal according to the chronological order of edges. 
%(lines 5-19)
Specifically, incoming edges are sampled in descending order of time, while outgoing edges are sampled in ascending order, until the predefined hop limit $k$ is reached.
Importantly, \texttt{fork} edges are exempt from the hop count to avoid evasion attacks that artificially extend the attack path by continuously forking.
This design generalizes the sampling across different operating systems and attack patterns.
During traversal, the dependencies of each visited node are inspected, with only those assessed suspicious being sampled. 
To effectively expose suspicious interactions concealed within numerous dependencies, we customized sampling rules for each node type. 
{\color{black}
As outlined in Table \ref{sampling rules}, these rules operate solely on abstracted behavioral semantics, i.e., dependency explosion flag (i.e., `exp') and abstract node type (i.e., `abs type') attributes in the PPG, rather than explicit attack signatures or OS-specific identifiers. This design ensures that the sampling logic generalizes across different platforms and previously unseen attack variants. 
}
%The design rationale is described below.
%by leveraging the dependency explosion flag (i.e., `exp') and abstract node type (i.e., `abs type') attributes available in the PPG.
%As outlined in Table \ref{sampling rules}, each rule utilizes the abstract node type (i.e., `abs type') and dependency explosion flag (i.e., `exp') attributes present in the PPG. 
%i.e., different traversed nodes are customized based on whether they are dependency explosion nodes(i.e., the `exp' flag is 1) or sparse nodes (i.e., the `exp' flag is 0), with the goal of covering the propagation paths of various suspicious information flows.
%Second, for each sampled node, especially the dependency explosion node, dependencies related to suspicious information flows are captured according to the sampling rules outlined in Table \ref{sampling rules}. 
%such as long-running processes (e.g., \texttt{firefox}) or system resource files (e.g., \texttt{ld-elf.so.hints}). These nodes have numerous dependencies and can dominate suspicious semantics if sampled blindly. Similarly, research that overlooks sampling these nodes  \cite{watson} or restricts the number of sampled nodes  \cite{megrapt} also risks losing critical semantic information. Our approach,
\begin{table*}[]
  \scriptsize
  \centering
  \caption{Interaction sampling rules. P, F and N denote set of corresponding abstract types of process, file, and netflow, respectively, with $v$, $u$, and $e$ representing the visited node, neighboring node, and edge.}
  \label{sampling rules}
  \begin{tabular}{
    |l|l|c|c|}
  \hline
  \textbf{Node} & \textbf{NO.} & \textbf{Sampling Events} & \textbf{Prerequisites} \\ \hline
  \multirow{7}{*}{Process} & R1 & \multirow{2}{*}{$e \in (Send,~Recv)~\wedge~ u.type \in \text{N}$} & $v.type \notin (web\_process) \wedge v.exp = 1$ \\ \cline{2-2} \cline{4-4} 
   & 
   R2 &  & $v.type \in \text{P} ~\wedge~v.exp = 0$ \\ \cline{2-4} 
   & 
   R3 & $e \in (*) ~\wedge ~ u.type \notin (unknown\_file)$ & \multirow{2}{*}{$v.type \in \text{P}~\wedge~v.exp = 0$} \\ \cline{2-3}
   & 
   R4 & $e \in (*)~\wedge~ u.type \in \text{P}$ &  \\ \cline{2-4} 
   & 
   R5 & $e \in (Modify,Write,Link,Rename)~\wedge~u.type \in (sys\_file,lib\_file)$ & \multirow{3}{*}{$ v.type \in \text{P}~\wedge~v.exp = 1 $} \\ \cline{2-3}
   & 
   R6 & $e \in (Modify)~\wedge~ u.type \in (sys\_process,serv\_process)$ &  \\ \cline{2-3}
   & 
   R7 & $e \in (*)~\wedge~u.type \in (cfg\_file)$ &  \\ \hline
  \multirow{2}{*}{File} & 
  R8 & $e \in (Modify,Write,Link,Rename) ~\wedge~u.type \in \text{P}$ & $ v.type \in (sys\_file,lib\_file,cfg\_file)~\wedge~v.exp = 1 $ \\ \cline{2-4} 
   & 
   R9 & $e \in (*)~\wedge~u.type \in \text{P}$ & $v.type \in \text{F}~\wedge~v.exp = 0 $ \\ \hline
  Netflow & 
  R10 & $e \in (Send,~Recv)~\wedge~u.type \in \text{P}$ & $v.type \in \text{N}$ \\ \hline
  \end{tabular}
  \end{table*}

\subsubsection{Sampling Rules} 
Since APT campaigns often entail communication with command and control (C\&C) servers and the exfiltration of sensitive data \cite{holmes}, we regard network interactions with public addresses as risk indicators. Based on this premise, we design rules \textbf{R1}, \textbf{R2} and \textbf{R10} for process and netflow nodes to selectively sample suspicious network interactions. 
Taking R10 as an illustrative example: if the visited node is a netflow node (i.e., $v \in \text{N}$), then network interactions with neighboring process nodes will be sampled (i.e., $e\in(Send, Recv)~\wedge~u\in \text{P}$). 
Similarly, R2 applies when the visited node is a process node. However, certain web applications (e.g., \texttt{Firefox.exe} in Figure \ref{challenge}b) routinely engage in extensive public network communications, which, if sampled without discrimination, could introduce substantial noise into the threat graph.
To mitigate this, R1 limits network interaction sampling when visiting a dependency explosion process (i.e., $v.exp = 1$) of the abstract type `web\_process'. 
\begin{comment}
\begin{equation*}
  e \in (*)~~\wedge~~s \in \text{P}, ~~\text{If } d \in \text{N}
\end{equation*}
where the 
\end{comment}
%Specifically, we assume the network information flows as suspicious, given that APT attacks often involve communication with command and control (C\&C) servers and sensitive data transfers  \cite{holmes,conan}, which generate network flows to and from public networks. This assumption prompts us to sample all network dependencies.
% linked with netflow nodes and process nodes.
It is worth noting that this restriction \textit{does not compromise} the overall attack semantics as C\&C communications are initiated not only through the browser but also extend to processes spawned by malicious payloads (e.g., `\texttt{gup.exe} $\to$ \texttt{53.192.68.50}' in Figure \ref{challenge}b), and these network interactions can be sampled by rules R1 and R10.

After compromising a system, attackers often work to solidify their foothold and advance their objectives (e.g., `Privilege Escalation' by process injection).
Following insights from \cite{holmes,aptshield} and the ATT\&CK framework \cite{attck}, we observe that such behaviors tend to generalize across various contexts, typically appearing as: 1) altering system or library files; 2) manipulating process environments, permissions, or resources; and 3) modifying configuration files (e.g., `\texttt{/etc/passwd}').
These information flows are considered high-risk, consistent with APTShield \cite{aptshield}.
To prevent noise, we customize two sets of sampling rules depending on whether the accessed file or process node exhibits dependency explosion. 
Specifically, for dependency explosion processes, interactions of modifying system or library files (\textbf{R5}), altering processes (\textbf{R6}), or configuration file operations (\textbf{R7}) will be sampled.
For sparse processes (i.e., $v.exp = 0$), \textbf{R3} and \textbf{R4} are applied to sample all interactions except those involving `unknown\_file', which usually represent low-risk files (e.g., logs files) and can be excluded from sampling. 
The sampling rules for file nodes mirror this logic: if the accessed file is sparse, all process interactions are sampled via \textbf{R9}; if it is a dependency explosion file of type `sys\_file', `lib\_file' or `cfg\_file', we apply \textbf{R8} to capture modifications to its content or properties.

These steps are executed recursively until all POIs are processed.
%(lines 2-21)
Dependencies among sampled nodes are then extracted from the PPG to form threat graphs.
%(line 20)
Finally, graphs with overlapping nodes are merged, nodes with identical names are consolidated, and the resulting graphs are sent to the attack representation and matching component.
%(lines 22-23)
%The resulting graphs are passed to the attack representation and matching component.
%And we perform graph simplify operations to stratify the threat graphs. First, graphs with overlapping POIs are first merged into one, and then the nodes with identical names in each threat graph are merged and the resulting graph is passed to the graph matching module.
%We also employ two strategies to simplify the graphs: first, merging threat graphs with overlapping POIs, and second, merging nodes with identical names within each threat graph.
%ProHunter's sampling algorithm runs on in-memory PPG, reducing performance delays caused by frequent disk accesses.
{\color{black}
\subsubsection{Complexity Analysis for Sampling.} 
This section discusses the computational complexity of the
threat graph sampling. Given a provenance graph $\mathcal{G}$ with POI set $\mathcal{S}$ and sampling hop $k$, naive BFS has complexity $\mathcal{O}(|\mathcal{S}| \cdot \bar{d}^k)$, where $\bar{d}$ is the average node degree. 
In real-world scenarios, $\bar{d}$ can be very large due to dependency explosion nodes, causing exponential growth.
Our heuristic sampling rules (Table \ref{sampling rules}) mitigate this through semantic pruning, and thus reduces the effective search branches to $\bar{d}_{e} \ll \bar{d}$, yielding complexity $\mathcal{O}(|\mathcal{S}| \cdot \bar{d}_{e}^k)$. With $|\mathcal{S}|$ and $k$ as small constants, and $\bar{d}_{e}$ bounded, sampling becomes independent of global graph size, thereby ensuring high efficiency.
}

\subsection{Attack Representation $\&$ Matching}
To encode the attack semantics of provenance graphs, existing methods primarily adopt uniform aggregation across localized behavior features of each node, referred to as \textit{static graph representation} \cite{basepaper}. 
%The core limitations of this method lie in its inability to accurately capture inherent attack semantics within threat graphs and its susceptibility to structural variations.
However, the structural inconsistencies between threat graphs and CTI-derived query graphs can lead to divergent representations due to semantic gaps or noise introduced by sampling (see Figure \ref{trace sampling graph} in Section \ref{sampling_algorithm_effectiveness}), adversely impairing threat hunting (Challenge C3).

\begin{figure}[]
  \centering
  \includegraphics[scale=0.65]{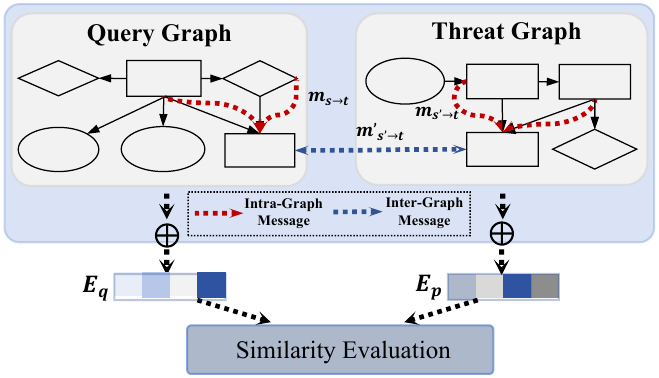}
  \caption{\centering{Illustration of adaptive graph representation.}}
  \label{graph matching framework}
\end{figure}

We posit that, despite these gaps, the core attack patterns preserved in both threat graphs and query graphs remain consistent. 
%Take the sampled threat graph of the E3-Cadets dataset as an example, as shown in Figure \ref{sample}. 
%The static graph representation scheme uniformly integrates the localized features of both noisy nodes (e.g., \texttt{procstart}, \texttt{random}) and suspicious nodes during the encoding process. This could result in graph embeddings that significantly diverge from the query graph when the sampled graph contains noisy nodes causing structural mutations, thereby affecting the hunting performance, corresponding to challenge C3.
%Compared to the query graph (i.e., black solid lines), if the sampled noisy node \texttt{procstat} induces significant structural changes in the sampled threat graph. This results in the static graph representation method generating two distinct graph embeddings that might impair the threat hunting performance, corresponding to challenge C3. \textsc{ProHunter} implements an adaptive graph representation model that introduces intra- and inter-graph message passing mechanisms to capture localized-globalized behavior patterns associations of matching graphs.
For instance, in Figure~\ref{challenge}, the process \texttt{gup.exe} in the threat graph contacts the C\&C server, loads the malicious payload \texttt{cKfGW.exe}, and spawns a related process, mirroring the behavior of \texttt{update.exe} in the query graph. A similar correspondence is observed for the \texttt{lsass.exe} process.
By mapping these established attack patterns in query graphs to their functionally equivalent in threat graphs, we can uncover their intrinsic semantic parallels.
Motivated by this insight, \textsc{ProHunter} integrates a feature enhancement strategy with an adaptive graph representation model to capture such shared attack patterns. First, it abstracts both threat and query graphs into a semantically aligned feature space. Then, as shown in Figure~\ref{graph matching framework}, it aggregates intra-graph messages ($m_{s \to t}$) from each node's neighbors to obtain localized behavior features. These features ($m_{s \to t}^{'}$) are further propagated across graphs via an inter-graph message passing mechanism, where an attention module prioritizes structurally and semantically similar patterns. By synthesizing both intra- and inter-graph contextual signals, \textsc{ProHunter} achieves robust alignment of localized and globalized attack semantics across diverse contexts.
%that uses GIN to extract localized behavior pattern features for each node by summarizing intra-graph messages passed by neighbor nodes. This is followed by an inter-graph message passing mechanism to learn the sub-behavior associations between matching graphs. By aggregating both intra-graph and inter-graph messages, this module captures the localized-globalized features between query graph and threat graph.

%ProHunter implements an adaptive graph representation model that introduces intra- and inter-graph message passing mechanisms to capture localized-globalized behavior patterns associations of matching graphs. 
%We employ the inter-graph message passing mechanism to learn the dominant local behavior patterns in each other's graphs to dig out the deep behavior pattern similarities, which are then combined with the globalized behavior pattern features to enhance resistance to the above inconsistencies.

\begin{comment}
\begin{figure}[!h]
  \centering
  \includegraphics[scale=0.7]{query_graph.pdf}
  \caption{\centering{\small{The attack `Nginx Backdoor with Drakon In-Memory' on 2018-04-13 of the E3-Cadets dataset. Black solid lines indicate the query graph. Red dashed lines represent sampling noise.}}}
  \label{sample}
\end{figure}
\end{comment}

{\color{black}
\subsubsection{Feature Initialization}
%Intra-graph messages are computed from the features of neighboring nodes within each graph and are subsequently aggregated to form globalized features. Inter-graph messages, on the other hand, are generated based on correlations in shared substructure features across matching graphs, with these messages being aggregated using attention mechanisms to create localized features. \textsc{ProHunter} then integrates these diverse feature types to produce comprehensive graph vectors.
As discussed in Section \ref{sec:semantic align}, using specific node names as graph features to extract attack representations preserves fine-grained details but is easily obfuscated, limiting generalizability in real-world threat hunting. Conversely, using node types as graph features is prone to false alarms when benign graphs exhibit similar type interactions to malicious ones \cite{megrapt}. 
Alternatively, \textsc{ProHunter} adopts intermediate-level abstracted node types in the PPG module (see Table \ref{absnodes}), which synthesizes semantic cues from both node types and node names into more generalized and informative attributes. This utilization is readily available without extra transformations.
Formally, each node in the query graph $G_q$ and the threat graph $G_p$ is encoded as a one-hot vector representing its abstract node type (e.g., $h_s$ for node $s$), while edges connecting node pairs are represented as multi-hot embeddings (e.g., $e_{st}$ for edges between nodes $s$ and $t$) capturing high-dimensional interaction features. 
} 

\subsubsection{Attack Representation}
After the initialization, intra-graph messages are generated by concatenating these features:

\begin{equation*}
  m_{s \to t} = (h_t || h_s || e_{st}),
  %, s \in \mathcal{N}(t).
\end{equation*}

\noindent where node $s$ is a neighbor of node $t$, and `$||$' denotes the concatenation operation. 
%The hidden embeddings of node $t$ and $s$ are represented by $h_t$ and $h_s$, respectively, while $e_{st}$ represents the embedding of the edges connecting nodes $s$ and $t$. 
The intra-graph messages from neighboring nodes are then aggregated to form the localized behavior feature of node $t$, followed by a ReLU activation to effectively capture complex interactions:

% Message Function Formula
\begin{equation*}
  h_{t}^{'} = ReLU ( W^{T}_{intra} \cdot \sum_{s \in \mathcal{N}(t)} m_{s \to t}).
\end{equation*}

After deriving localized behavior features for both threat and query graphs, an inter-graph message passing mechanism \cite{neuralmatch1} is employed to exchange these features between them, thereby capturing their analogous attack patterns. The inter-graph message is defined as follows:

% Message Function Formula
\begin{equation*}
  m_{s \to t}^{'} = a_{s \to t} \cdot h_{s}^{'},
  %\text{ where } s \in \mathcal{G}^{'}, t \in \mathcal{G}.
\end{equation*}
%\begin{equation*}
%  a_{s \to t} = Sim(h_{s},h_{t}), s \in \mathcal{G}^{'}.
%\end{equation*}

\noindent where nodes $s$ and $t$ originate from the threat graph and query graph (or vice versa), respectively. Here, $h_{s}^{'}$ represents the behavior feature of node $s$, and $a_{s \to t}$ denotes the attention weight calculated using the \textit{dot product similarity} between $h_{s}^{'}$ and $h_{t}^{'}$. This attention-based mechanism ensures that analogous patterns receive higher emphasis, thus facilitating adaptive matching across diverse threat and query graphs.

To mitigate computational burden, feature exchanges are limited to nodes with degrees greater than 3, as these nodes exhibit dominant behaviors and encapsulate richer semantics, thus ensuring effective semantic alignment.
%This strategy reduces computational overhead and ensures effective semantic alignment.
Upon receiving inter-graph messages, we update the feature of node $t$:

% Message Function Formula
\begin{equation*}
  h_{t}^{*} = MLP ( h_{t}^{'} || \sum_{s \in \mathcal{G}^{'}} m_{s \to t}^{'}).  
\end{equation*}

\noindent where intra-graph and inter-graph messages are aggregated using a two-layer MLP, resulting in the updated $h_{t}^{*}$. By recursively applying this procedure, the interaction semantics of a broader context can be effectively captured.
Finally, we apply summation pooling over all node features in each query and threat graph to derive the robust graph representations $E_q$ and $E_s$. These representations effectively encapsulate both localized and globalized similarities in attack semantics.
%Finally, the cosine similarity between $E_q$ and $E_s$ are calculated as the threat score for risk assessment. By comparing the threat score against a predefined threat threshold $\theta$, we can determine whether the sampled threat graph is malicious.

\subsubsection{Model Training} \label{training}
To enhance the model's discriminative power in graph matching, we employ contrastive learning that optimizes representations by learning similar embeddings for behaviorally analogous graphs while ensuring distinct embeddings for dissimilar pairs via contrastive loss minimization.

Note that the graph matching model is designed to predict behavior similarities of graph pairs, rather than performing binary classification of benign versus malicious, which allows the use of benign datasets to construct the training set. 
To this end, we extract benign audit logs from each DARPA dataset recorded before any attack phase, and transform them into provenance graphs.
From these graphs, we sample subgraphs using a BFS strategy with hop sizes ranging from 2 to 4, while constraining each sample to contain between 10 and 30 nodes.
This sampling policy ensures that the structural characteristics of the sampled training graphs align with those of real-world scenarios, thereby ensuring better generalization.
In total, 15,000 sampled graphs are generated, denoted as $\mathcal{G}_{ori}$. 
By excluding any data from the attack execution phase, we ensure that the threat graphs used in testing remain entirely unseen during training, guaranteeing a fair and reliable evaluation.

%We randomly sample nodes of degree greater than 5 from the benign dataset and perform a random BFS. The sampling depth is limited to 3, and the number of sampled nodes is limited to between 10 and 30 to align with the threat hunting scenario  \cite{poirot,provgsearcher}. In total, 10000 graphs are sampled for each dataset, denoted as $\mathcal{G}_{raw}$.
%Since the graphs in $\mathcal{G}_{raw}$ available for forming positive graph pairs (i.e. similar graphs) may be limited, 
As previously mentioned, contrastive learning relies on positive and negative graph pairs as supervised signals. However, ensuring that randomly sampled graphs $\mathcal{G}_{ori}$ contain sufficient positive pairs poses a significant challenge. 
To mitigate this issue, we apply two graph augmentation strategies: edge perturbation and node perturbation~\cite{aug}. These strategies construct augmented graphs with analogous behavior patterns for each graph in $\mathcal{G}_{ori}$, yielding $\mathcal{G}_{pos}$, a refined set of positive graph pairs.
Details of these augmentation strategies are provided in \ref{graph augmentation}.
%1) \textbf{Edge Perturbation.} Randomly adds or removes edges of 'Process $\to$ File' or 'Process $\to$ NetFlow', as defined in Table \ref{nodes edges}. 2) \textbf{Node Perturbation.} Randomly removes `File' or `NetFlow' nodes or incorporates nodes and edges from other graphs within $\mathcal{G}_{raw}$.
%Specifically, the sampler selects nodes with degrees greater than 5 and performs random BFS with a maximum hop of 3. The number of sampled nodes is constrained to between 10 and 30 to match the threat hunting scenario, which is same as  \cite{poirot,provgsearcher}. In total, 10,000 graphs are generated for each dataset, referred to as $\mathcal{G}_{original}$.
%we first sample subgraphs or nodes to perturb edges and nodes, creating variations that retain the core structural properties of the original graphs
%Specifically, edge perturbation method randomly adds edges of $'process \to file'$ and $'process \to network'$ (defined in Table \ref{nodes edges}) to the graphs. Node perturbation, on the other hand, randomly incorporates nodes and corresponding edges from other graphs in $\mathcal{G}_{raw}$ into the augmented graph. 
%The perturbation ratio is set to 20\%, thereby creating positive sample pairs $\mathcal{G}_{pos}$ for each dataset.
For negative graph pairs, a naïve approach would be to randomly select graph pairs from $\mathcal{G}_{ori}$. However, such random selections may inadvertently group semantically similar graphs, potentially diminishing the model's learning efficacy \cite{contrastive}.
To address this issue, we identify negative pairs with notable structural differences within $\mathcal{G}_{ori}$ using the Graph Edit Distance (GED) algorithm \cite{bipartite}, which quantifies the structural similarity by calculating the minimum number of edits required to transform one graph into another. 
Specifically, for each graph in $\mathcal{G}_{ori}$, we select another graph whose GED exceeds $min(|\mathcal{V}_a|+|\mathcal{E}_a|,|\mathcal{V}_b|+|\mathcal{E}_b|)$ %\footnote{$|V|$ and $|E|$ denote the number of nodes and edges in the graph.}
 to form negative pairs. 
This process generates $\mathcal{G}_{neg}$, a refined set of negative pairs.
Once $\mathcal{G}_{pos}$ and $\mathcal{G}_{neg}$ are generated, we define the following contrastive loss to optimize the model: 
%and achieve our objectives:

\begin{equation*}
\mathcal{L} = - \frac{1}{|\mathcal{G}_{pos}|} \sum_{(\mathcal{G}_{i},\mathcal{G}_{j}) \in \mathcal{G}_{pos}} \log \left( \frac{e^{\mathrm{sim}({\mathcal{G}_{i},\mathcal{G}_{j})}/\tau}}{\sum_{(\mathcal{G}_{i},\mathcal{G}_{k}) \in \mathcal{G}_{neg}} e^{\mathrm{sim}{(\mathcal{G}_{i},\mathcal{G}_{k})}/\tau}} \right),
\end{equation*}

\noindent where $\mathrm{sim}(\mathcal{G}_{i},\mathcal{G}_{j})$ denotes the cosine similarity score computed from attack representations for each positive pair $(\mathcal{G}_{i},\mathcal{G}_{j})$ in $\mathcal{G}_{pos}$. The denominator sums similarity scores for all negative pairs $(\mathcal{G}_{i},*)$ in $\mathcal{G}_{neg}$. The temperature parameter $\tau$ scales these scores.
This contrastive learning framework trains the model to produce higher similarity scores for behaviorally similar graphs and lower scores for dissimilar ones.

\subsubsection{Model Inference} \label{testing}
During threat hunting, \textsc{ProHunter} takes each sampled threat graph and CTI-derived query graph as input, generating attack representations $E_q$ and $E_p$. It then computes the cosine similarity between them to produce a threat score for risk assessment. By comparing this score against a predefined threshold $\theta$, \textsc{ProHunter} determines whether the system is compromised.

{\color{black}
\subsubsection{Complexity Analysis for Representation.} 
This section derives the computational complexity of the graph representation module.
Let \(|\mathcal{V}_p|\), \(|\mathcal{E}_p|\) denote the number of nodes and edges in a threat graph, \(|\mathcal{V}_q|\), \(|\mathcal{E}_q|\) for a query graph, \(d\) be the feature dimension, and \(L\) be the number of message passing layers. 
For each graph, intra-graph message passing requires \(O(|\mathcal{E}| \cdot d)\) for message computation and \(O(|\mathcal{V}| \cdot d^2)\) for feature updates. Therefore, the total cost of intra-graph propagation is: 
\(O(L \cdot ((|\mathcal{E}_p| + |\mathcal{E}_q|) \cdot d + (|\mathcal{V}_p| + |\mathcal{V}_q|) \cdot d^2)).\)
For inter-graph message passing, cross-graph attention costs \(O(L \cdot |\mathcal{V}_p^{'}| \cdot |\mathcal{V}_q^{'}| \cdot d)\), where \(|\mathcal{V}_p^{'}|\) and \(|\mathcal{V}_q^{'}|\) denote the number of high-degree nodes selected from each graph for local semantic comparison, respectively. Combining both parts, the overall complexity is: 
\(O(L \cdot ((|\mathcal{E}_p| + |\mathcal{E}_q|) \cdot d + |\mathcal{V}_p^{'}| \cdot |\mathcal{V}_q^{'}| \cdot d + (|\mathcal{V}_p| + |\mathcal{V}_q|) \cdot d^2)).\)
}
%sim(\mathcal{G}_{i},\mathcal{G}_{k})
 
\section{Evaluation}

{
\color{black}
We conducted an extensive evaluation of \textsc{ProHunter} to assess its performance across diverse attack scenarios.
Section~\ref{sec:settings} first outlines the experimental setup.
We then examine the performance of individual components and explore the system's adaptability to different POI categories in Section~\ref{effectiveness}.
An ablation study in Section~\ref{abl} isolates the contribution of our attack representation design, while Section~\ref{comp exp} benchmarks \textsc{ProHunter} against state-of-the-art approaches in memory compaction, threat graph sampling, and threat hunting.
A detailed breakdown of system overhead is presented in Section~\ref{sec:overhead}.
Due to space constraints, %detailed descriptions of the experimental environment and implementation settings are provided in Appendix~\ref{app:env}.Appendix \ref{case study} provides a case study of threat hunting with \textsc{ProHunter}, while Appendix \ref{app:tuning exp} elaborates on parameter tuning experiments.
a real-world case study, and parameter tuning results are provided in \ref{case study} and \ref{app:tuning exp}.
}

\subsection{Experiment Settings} \label{sec:settings}

\subsubsection{Evaluation Datasets}\label{sec:dataset}
{\color{black}
We evaluate \textsc{ProHunter} using datasets from the DARPA Transparent Computing Program, including Engagement 3 (E3) \cite{darpatce3}, Engagement 5 (E5) \cite{darpatce5} and Operationally Transparent Cyber (OpTC) \cite{darpatcoptc}, which cover multiple APT scenarios across distinct operating systems. 
These datasets originate from red-team vs. black-team exercises simulating real-world APTs within enterprise networks, where benign activities (e.g., web browsing, email usage, SSH logins) are recorded alongside attack traces, enabling rigorous cross-platform assessment.
The E3 project lasts for a fortnight, with benign activities logged in the first week and attacks executed in the second.
The E5 project lasts nine days, with attacks confined to specific days between 9 AM and 5 PM.
We selected audit logs from Cadets, Theia, Trace and Clearscope for their detailed after-action reports and rich event records. These datasets cover Linux, FreeBSD, and Android environments.
We also utilized the OpTC dataset, which originates from the most recent DARPA TC iteration \cite{optcanalysis}. This large-scale dataset was captured over two weeks from an enterprise network of 1,000 Windows hosts, accumulating over 17 billion events. 
From OpTC, which includes ``Malicious Upgrade'', ``Custom Powershell Empire'', and ``Plain PowerShell Empire'' attack campaigns, we utilized both the benign logs and the attack logs from compromised hosts as our evaluation data.

To assess PPG compaction and adaptability for practical use, we employ audit logs of varying durations (1–10 days), summarized in Table \ref{detail_datasets}. For threat hunting evaluation, query graphs are first generated from after-action reports \cite{darpatce3,darpatcoptc} using the method in Section~\ref{query generate}. The statistics of query graphs are presented in Table \ref{sampling analysis}. 
%We labeled genuine threat graphs from all sampled graphs to verify hunting accuracy based on attack occurrence times recorded in after-action reports, and those time-mismatched graphs were paired with query graphs to create negative samples for evaluating the effectiveness of false alarm reduction.
Then, a set of POIs is identified and labeled as false alarms or true positives by cross-checking their initialized timestamps and entity names against the ground truth in after-action reports, consistent with \cite{megrapt}. Our heuristic sampling algorithm then extracts 3,846 graphs from the provenance graphs, with about 0.7\% being threat graphs. These graphs are utilized to evaluate the sampling algorithm and the threat hunting performance in distinguishing false alarms from true positives across various platforms.
%Next, threat graphs are sampled based on identified POIs using the algorithm delineated in Section \ref{sec:sampling}, and their ground truth labels are manually annotated in accordance with sampled timestamps and attack occurrence times documented in after-action reports.
%The threat graphs are then compared with their matched query graphs to assess the precision of sampling anomalous interactions. 
%Finally, these annotated threat graphs are combined with query graphs to form positive and negative sample pairs for threat hunting evaluation.

\begin{table}[t]
  \scriptsize
  \centering
  \caption{\centering{\textcolor{black}{Log files used in compaction evaluation.}}}
  \label{detail_datasets}
  \setlength{\tabcolsep}{4pt} % 减少列间距
  \begin{tabular}{|l|l|l|}
  \hline
  {\color{black} \textbf{Dataset}} & {\color{black} \textbf{Log Files}}   &  {\color{black} \textbf{Platform} }
  \\  \hline  \multicolumn{1}{c}{} \vspace{-6.5pt} \\\hline
  {\color{black} E3-Cadets} & {\color{black} ta1-cadets-e3-official-1.json.\{0-4\}} & {\color{black}FreeBSD} \\ \hline
  {\color{black} E3-Theia} & {\color{black} ta1-theia-e3-official-6r.json.(0-12)} & {\color{black}Linux} \\ \hline
  {\color{black} E3-Trace} & {\color{black} ta1-trace-e3-official.json.\{0-203\}} & {\color{black}Linux} \\ \hline
  {\color{black} E5-Theia} & {\color{black} ta1-theia-1-e5-official-2.bin.\{27-31\}} & {\color{black}Linux} \\ \hline
  {\color{black} E5-Clearscope2\_1} & {\color{black} ta1-clearscope-2-e5-official-1.bin.\{15-20\}} & {\color{black}Android} \\ \hline
  {\color{black} E5-Clearscope2\_2} & {\color{black} ta1-clearscope-2-e5-official-1.bin.\{24-33\}} & {\color{black}Android} \\ \hline
  {\color{black} OPTC} & {\color{black} benign/20-23Sep19/AIA-201-225/\{*\}} & {\color{black}Windows} \\ \hline
  \end{tabular}
\end{table}
\subsubsection{Evaluation Metrics}\label{sec:metric}
To evaluate the compaction capability, we use main memory consumption as the metric. 
{\color{black}
For the threat graph sampling algorithm, we assess its recall and precision performance through customized indicators of Noise Rate (NR) and Coverage Rate (CR):
}
%both of which are crucial for security analysts to conduct quality investigation. Using the interactions in the query graph as ground truth, the formulas are 

{\footnotesize
\begin{equation*}
  NR = \frac{1}{N} \sum_{k=1}^{N} \frac{|\mathcal{SG}_{k} - \mathcal{QG}_{k}|}{|\mathcal{SG}_{k}|}, \quad CR = \frac{1}{N} \sum_{k=1}^{N} \frac{|\mathcal{SG}_{k} \cap \mathcal{QG}_{k}|}{|\mathcal{QG}_{k}|},
  \end{equation*}
} 
\begin{comment}
    {\footnotesize
  \begin{equation*}
  Coverage = \frac{1}{N} \sum_{k=1}^{N} \frac{|\mathcal{SG}_{k} \cap \mathcal{QG}_{k}|}{|\mathcal{QG}_{k}|}.
  \end{equation*}
  } 
\end{comment}

\noindent where $N$ is the number of query graphs, $\mathcal{QG}_k$ denotes the $k$-th query graph, and $\mathcal{SG}_k$ refers to the sampled threat graph matched with $\mathcal{QG}_k$. The coverage rate reflects the extent to which anomalous interactions are retained, while the noise rate quantifies the proportion of extraneous interactions captured.
%The coverage metric quantifies the overlap in nodes and edges between the matched threat graph and query graph, assessing the precision of sampling anomalous interactions. 
%In contrast, the noise ratio metric measures the proportion of nodes and edges in the threat graph that are missing from the matched query graph.
Threat hunting performance is evaluated using Accuracy, False Positive Rate (FPR), Recall, and Area Under the ROC Curve (AUC), consistent with prior work \cite{provgsearcher, megrapt, poirot}.
}

\begin{table*}[t]
  \scriptsize
  \centering
  \caption{\centering{Memory usage of PPG for storing audit logs of varying durations across different datasets.}}
  \label{ppg analysis}
  \setlength{\tabcolsep}{6pt} % 减少列间距
  \begin{tabular}{|l|l|l|c|c|c|c|c|}
    \hline
    \textbf{Dataset} &
    \textbf{\begin{tabular}[c]{@{}c@{}}Start Time\\ (YYYY-mm-dd \\ HH:MM:SS)\end{tabular}} &
    \textbf{\begin{tabular}[c]{@{}c@{}}End Time\\ (YYYY-mm-dd \\ HH:MM:SS)\end{tabular}} &
    \textbf{\begin{tabular}[c]{@{}c@{}}Duration\\ (HH:MM:SS)\end{tabular}} &
    \textbf{\#Entities} &
    \textbf{\begin{tabular}[c]{@{}c@{}}\#Events\end{tabular}} &
    \textbf{\begin{tabular}[c]{@{}c@{}}Total Memory \\(MB)\end{tabular}} &
    \textbf{\begin{tabular}[c]{@{}c@{}}Daily Memory \\(MB/Day)\end{tabular}}
    \\  \hline  \multicolumn{1}{c}{} \vspace{-6.5pt} \\\hline
    E3-Cadets &
      \begin{tabular}[c]{@{}c@{}}2018/4/6 14:01:15\end{tabular} &
      \begin{tabular}[c]{@{}c@{}}2018/4/11 15:16:13\end{tabular} &
      121:14:58 &
      259K &
      2.3M &
      17.27 &
      3.42 \\\hline
    E3-Theia &
      \begin{tabular}[c]{@{}c@{}}2018/4/10 12:44:33\end{tabular} &
      \begin{tabular}[c]{@{}c@{}}2018/4/13  17:04:30\end{tabular} &
      76:19:57 &
      823K &
      4.5M &
      37.12 &
      11.71  \\\hline
    E3-Trace &
      \begin{tabular}[c]{@{}c@{}}2018/4/2  17:14:04\end{tabular} &
      \begin{tabular}[c]{@{}c@{}}2018/4/13  9:06:49\end{tabular} &
      255:52:45 &
      6M &
      12.2M &
      184.02 &
      17.28 \\\hline
      {\color{black}E5-Theia} & {\color{black}\begin{tabular}[c]{@{}c@{}}2019/5/14 17:33:36\end{tabular}} & {\color{black}\begin{tabular}[c]{@{}c@{}}2019/5/16  7:40:22\end{tabular}} & {\color{black}38:06:46} & {\color{black}287K} & {\color{black}6.2M} & {\color{black}40.10} & {\color{black}25.26} \\\hline
      {\color{black}E5-Clearscope1} & {\color{black}\begin{tabular}[c]{@{}c@{}}2019/5/13 22:55:41\end{tabular}} & {\color{black}\begin{tabular}[c]{@{}c@{}}2019/5/15  18:00:06\end{tabular}} & {\color{black}43:04:25} & {\color{black}49K} & {\color{black}2.1M} & {\color{black}6.21} & {\color{black}3.47} \\\hline
      {\color{black}E5-Clearscope2} & {\color{black}\begin{tabular}[c]{@{}c@{}}2019/5/16  10:40:20\end{tabular}} & {\color{black}\begin{tabular}[c]{@{}c@{}}2019/5/20  5:53:51\end{tabular}} & {\color{black}91:13:31} & {\color{black}59K} & {\color{black}4.5M} & {\color{black}11.25} & {\color{black}2.97}\\\hline
    OPTC &
      \begin{tabular}[c]{@{}c@{}}2019/9/20 5:11:39\end{tabular} &
      \begin{tabular}[c]{@{}c@{}}2019/9/23 21:05:50\end{tabular} &
      87:54:10 &
      2M &
      4M &
      67.05 &
      18.37\\ \hline
    \end{tabular}
  \end{table*}

\subsubsection{Environment and Implementation}
The PPG module and threat graph sampling algorithm were evaluated on a Windows 10 machine with an Intel Core i5-11400 @ 2.60GHz CPU and 48 GB of RAM, with the implementation consisting of approximately 4,300 lines of C++ code, utilizing the `nlohmann/json' library  \cite{json} and Microsoft Visual C++ compiler  \cite{msvc}.
The attack representation and matching component were evaluated on an Ubuntu 20.04 system featuring an Intel Xeon Gold 5128 CPU, an NVIDIA RTX 4090 GPU, and 128 GB of RAM. This implementation comprises about 5,000 lines of Python code, developed using PyTorch  \cite{pytorch},
%\footnote{https://pytorch.org/.}
 NetworkX \cite{networkx},
 %\footnote{https://networkx.org/.} 
 and PyG \cite{pyg}. 
 %\footnote{https://www.pyg.org/.} 
{\color{black}
\subsubsection{Parameter Settings}
For the PPG component, we apply the default bit settings specified in Section \ref{sec:ppg}.
The sampling hop $k$ in the sampling algorithm is set to 2-3, which enables to sample an attack path with length of 5-7 in the provenance graph.
In the attack representation and matching component, we set the threat threshold $\theta = 0.3$ (alternatives discussed in Section \ref{threshold flexibility}).
To balance overhead and efficacy of threat hunting, the embedding size and message-passing layer are configured as $d = 128$ and $l = 3$.
The model is trained for 100 epochs with batch size 16, using Adam optimizer with learning rate 0.001.
For contrastive learning loss, we apply temperature $\tau = 0.1$.
Detailed hyperparameter sensitivity analysis is provided in \ref{app:tuning exp}.
}

\begin{table*}[!t]
  \scriptsize
  \centering
  \caption{{\color{black}\centering{Comparison of sampled threat graphs and ground-truth query graphs: Coverage Rate (CR) and Noise Rate (NR) analysis.}}}
  \label{sampling analysis}
  \setlength{\tabcolsep}{3.7pt} % 减少列间距
  \begin{tabular}{
    |l|    c|    c|    c|    c|    c|    c|    c|    c|    c| }
  \hline
  \multirow{2}{*}{\textbf{Dataset}} & \multirow{2}{*}{\textbf{\begin{tabular}[c]{@{}c@{}}\# Query Graphs\end{tabular}}} & \multicolumn{2}{c|}{\textbf{Query Graphs}}                                                                                                                  & \multicolumn{2}{c|}{\textbf{Threat Graphs}}                                                                                                                             & \multicolumn{2}{c|}{\textbf{CR}}             & \multicolumn{2}{c|}{\textbf{NR  }}           \\ \cline{3-10} 
                                    &                                                                                     & \multicolumn{1}{c|}{\textbf{\begin{tabular}[c]{@{}c@{}}Avg. \#  Nodes\end{tabular}}} & \textbf{\begin{tabular}[c]{@{}c@{}}Avg. \# Edges\end{tabular}} & \multicolumn{1}{c|}{\textbf{\begin{tabular}[c]{@{}c@{}}Avg. \#Nodes (ambiguous)\end{tabular}}} & \textbf{\begin{tabular}[c]{@{}c@{}}Avg. \#Edges (ambiguous)\end{tabular}} & \multicolumn{1}{c|}{\textbf{Node}} & \textbf{Edge} & \multicolumn{1}{c|}{\textbf{Node}} & \textbf{Edge} 
\\  \hline  \multicolumn{1}{c}{} \vspace{-6.5pt} \\\hline
E3-Cadets                            & 4                                                                                                               & \multicolumn{1}{c|}{19}                                                                & 31                                                                & \multicolumn{1}{c|}{14 (2)}                                                                   & 21 (3)                                                                   & \multicolumn{1}{c|}{0.77}          & 0.74          & \multicolumn{1}{c|}{0.10}           & 0.17          \\\hline
E3-Theia                             & 5                                                                                                                & \multicolumn{1}{c|}{12}                                                                & 18                                                                & \multicolumn{1}{c|}{15 (2)}                                                                   & 27 (4)                                                                   & \multicolumn{1}{c|}{0.73}          & 0.75          & \multicolumn{1}{c|}{0.19}          & 0.25         \\\hline
E3-Trace                             & 2                                                                                                               & \multicolumn{1}{c|}{12}                                                                & 20                                                                & \multicolumn{1}{c|}{23 (12)}                                                                  & 35 (16)                                                                  & \multicolumn{1}{c|}{0.79}          & 0.80           & \multicolumn{1}{c|}{0.60}           & 0.57         \\\hline
{\color{black} E5-Theia} & {\color{black} 2} & \multicolumn{1}{c|}{{\color{black} 10}} & {\color{black} 15} & \multicolumn{1}{c|}{{\color{black} 14 (3)}} & {\color{black} 19 (3)} & \multicolumn{1}{c|}{{\color{black} 0.90}} & {\color{black} 0.87} & \multicolumn{1}{c|}{{\color{black} 0.30}} & {\color{black} 0.27} \\ \hline
{\color{black} E5-Clearscope} & {\color{black} 2} & \multicolumn{1}{c|}{{\color{black} 8}} & {\color{black} 12} & \multicolumn{1}{c|}{{\color{black} 7 (0)}} & {\color{black} 10 (0)} & \multicolumn{1}{c|}{{\color{black} 0.88}} & {\color{black} 0.83} & \multicolumn{1}{c|}{{\color{black} 0.00}} & {\color{black} 0.00} \\ \hline
  OpTC                              & 15                                                                                                               & \multicolumn{1}{c|}{15}                                                                & 17                                                                & \multicolumn{1}{c|}{16 (2)}                                                                   & 21 (6)                                                                   & \multicolumn{1}{c|}{0.89}          & 0.88          & \multicolumn{1}{c|}{0.22}          & 0.28          \\ \hline
  \end{tabular}
  \end{table*}

\subsection{Effectiveness Analysis} \label{effectiveness}
\subsubsection{Graph Compaction Effectiveness}
We evaluate PPG's storage efficiency on provenance graphs collected over multiple days from operating systems, including Windows, Linux, FreeBSD, and Android.
These graphs are constructed by parsing deduplicated batches of audit logs, as detailed in \ref{generic_compression}.
%Two generic lossless compression strategies are implemented: consecutive duplicate events reduction ($S1$), time window-based duplicate network events reduction ($S2$).

% \usepackage{multirow}
As illustrated in Table \ref{ppg analysis}, the number of entities and interactions remain within the same order of magnitude (i.e., 2x to 9x) after event deduplication.
%which illustrates that the compression strategies of $S1$ and $S2$ exhibits varying degrees of effectiveness across different datasets. The $S1$ strategy significantly compress the E3 datasets, achieving compression ratios of 60\% to 90\%. This indicates that the audit tracking tools in E3 ensure the integrity of audit events. Second, the compression ratio of $S2$ in OpTC is 50\% to 60\%, reflecting a significant reduction in network events. This can be attributed to the large-scale network environment of the OpTC (1000 hosts). 
% without employing reduction strategies  \cite{kellect}. 
%The compression of consecutive duplicate events in OpTC is less than 2\%, which may be due to audit frameworks differences and preliminary event reduction strategies implemented by OpTC. 
%Third, the node versioning-based reduction strategy achieves a compression ratio of around 20\% for redundant information in the remaining events. Specifically, the compression ratios for the Cadets and Theia datasets reach 42\% and 29\%, respectively.
{\color{black} Moreover, the PPG maintains provenance graphs with daily memory usage ranging from 3 to 25 MB across various platforms. 
In the E3-Cadets and E5-Clearscope datasets, daily usage is only about 3 MB, owing to the fewer system entities and dependencies, which results in fewer extended structures being required.
By contrast, the E5-Theia dataset, with its massive interaction records, incurs a memory overhead of nearly 25 MB per day, yet still remains well within enterprise memory budgets \cite{arewethere}. Constructing provenance graphs for up to 90 days requires at most 2.2 GB of memory, enabling efficient analysis of the complete APT lifecycle.}
Interestingly, the amount of entities and dependencies does not scale linearly with memory usage, potentially due to varying distributions of sparse nodes and dependency explosion nodes across different datasets.
%degrees of compression applied to events by the node versioning control mechanism.
Overall, \textsc{ProHunter} constructs the provenance graph with low memory overhead and ensures real-time APT analysis.
%The PPG storage scheme ensures real-time access to the provenance graph and meets industry requirements for daily memory overhead \cite{arewethere}. 

%In the next subsection, we verify whether PPG's compression of the origin graph loses contextual semantic information between key entities by sampling the subgraphs of PPG and comparing them with the query graph.
\begin{figure}[!t]
  \centering
  \includegraphics[scale=0.7]{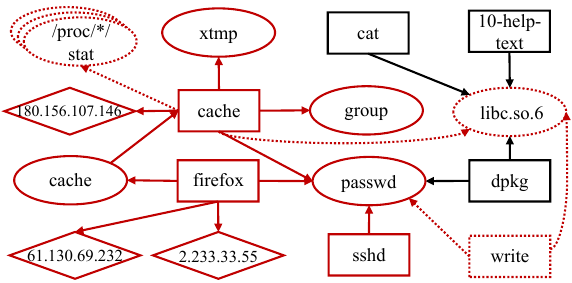}
  \caption{Threat graph sampled from E3-Trace: the attack `Firefox Backdoor with Drakon In-Memory'.}
  \label{trace sampling graph}
\end{figure}
\subsubsection{Sampling Algorithm Effectiveness} \label{sampling_algorithm_effectiveness}
We evaluate the effectiveness of our sampling algorithm by comparing sampled threat graphs from each dataset with their matched query graphs, using coverage and noise rate as indicators.
\begin{comment}
\begin{figure}[!t]
  \centering
  \includegraphics[width=3in]{samp_eff.pdf}
  \caption{\textcolor{black}{An attack graph from the E5-Clearscope. Red nodes indicate attack-related entities, solid nodes indicate sampled by \textsc{ProHunter}, and dashed nodes are missed attack nodes.}}
  \label{samp_eff}
\end{figure}
\end{comment}
{\color{black}
The results presented in Table \ref{sampling analysis} reveal that the average threat graph sizes closely match those of the query graphs.
Malicious nodes coverage consistently ranges from 70\% to 90\% (missing only 2 or 3 nodes), and edge coverage correlates positively with nodes. These results
indicate that \textsc{ProHunter} effectively extracts suspicious interactions, and the PPG component well-preserves critical dependencies.
%Figure \ref{samp_eff} illustrates a threat graph sampled by \textsc{ProHunter} in E5-Clearscope, which effectively retains the core semantics of suspicious behaviors.
The last column reveals that noise introduced is approximately 20\% (with only 2 to 3 noisy nodes sampled), and notably, none at all in E5-Clearscope.
}

We observe that the noise is higher in E3-Trace due to certain ambiguous behaviors, potentially malicious but absent from after-action reports, being counted as noise. 
Figure \ref{trace sampling graph} highlights one such case: the payload \texttt{cache} loads the modified library \texttt{lib.so.6} and accesses process data in \texttt{/proc}. These activities (dashed nodes and edges), though not documented, may indicate ongoing compromise.
The findings validate that our sampling algorithm captures not only critical attack interactions but also suspicious activities overlooked by CTIs, offering security analysts valuable insights for threat analysis.
We summarize these uncertain quantities in the `(ambiguous)' column of Table \ref{sampling analysis}.
{\color{black}
These results collectively demonstrate that the proposed sampling rules preserve core attack semantics with high recall while keeping noise at a manageable level.}
%In summary, our threat graph sampling algorithm enhances the effectiveness of threat hunting, and empowers security analysts with actionable insights for more impactful assessments. 
%Although 2 to 3 malicious entities may be omitted, the remaining core attack semantics still enable faster and more focused assessments. 

% \usepackage{multirow}

\begin{figure*}[t]
  \centering
  \subfloat[\small{{\color{black}Anomaly score distribution w/ suspicious nodes as POIs.}}]{\includegraphics[width=2.34in]{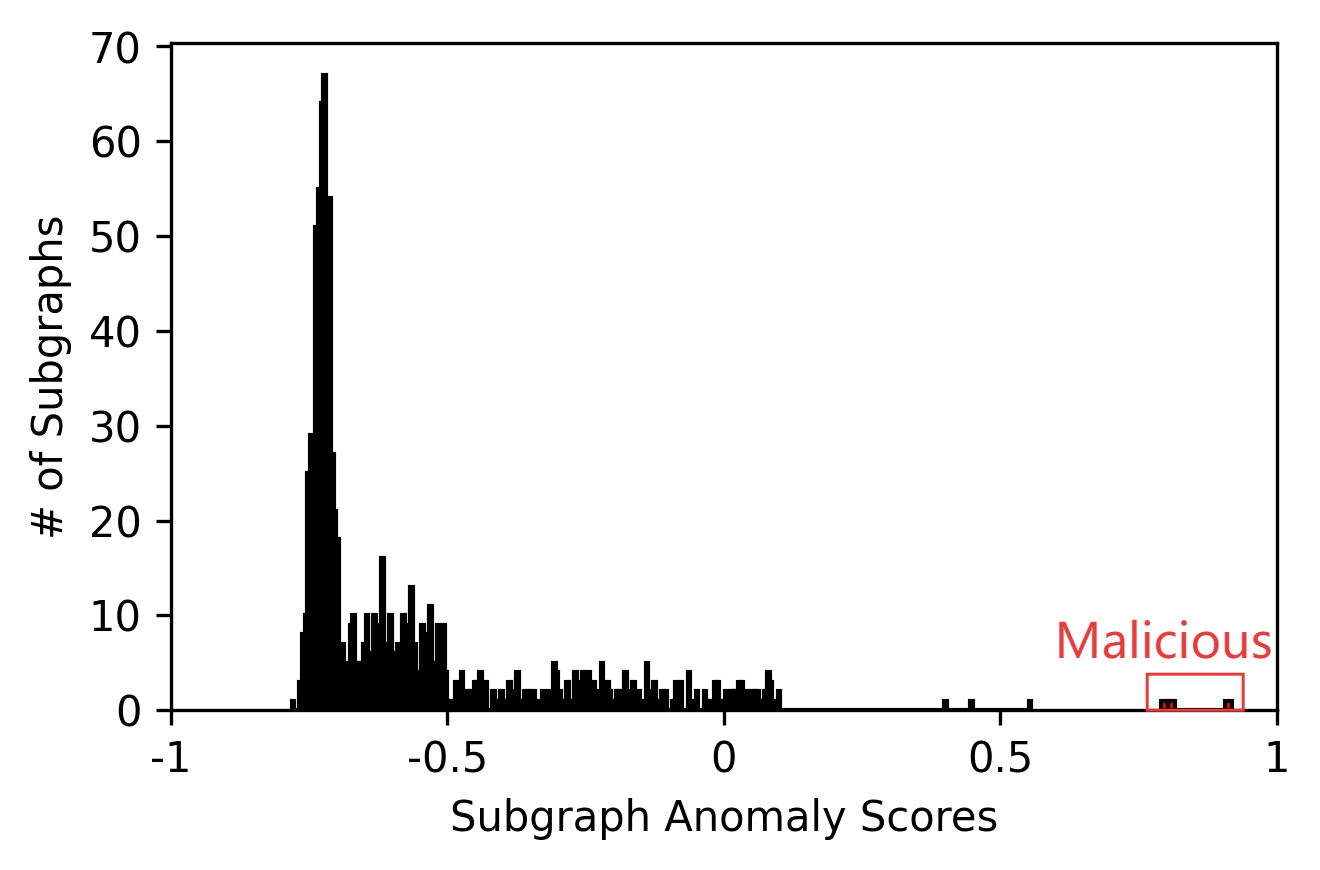}%
  \label{poi quality}}
  \hfil
  \subfloat[\small{{\color{black}Anomaly score distribution w/o POIs.}}]{\includegraphics[width=2.3in]{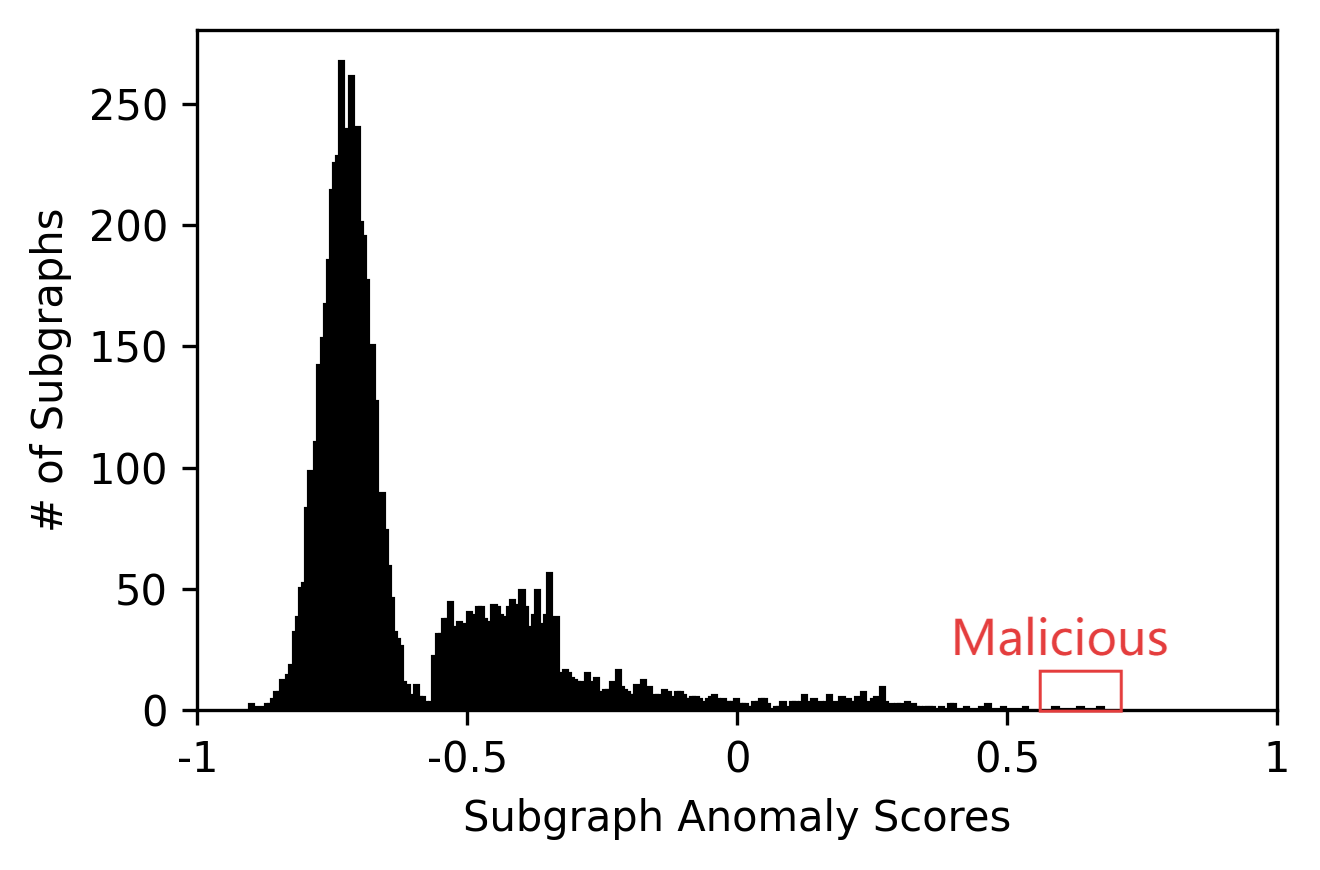}%
  \label{without poi}}
  \hfil
  \subfloat[\small{{\color{black}Time cost distribution for sampling.}}]{\includegraphics[width=2.3in]{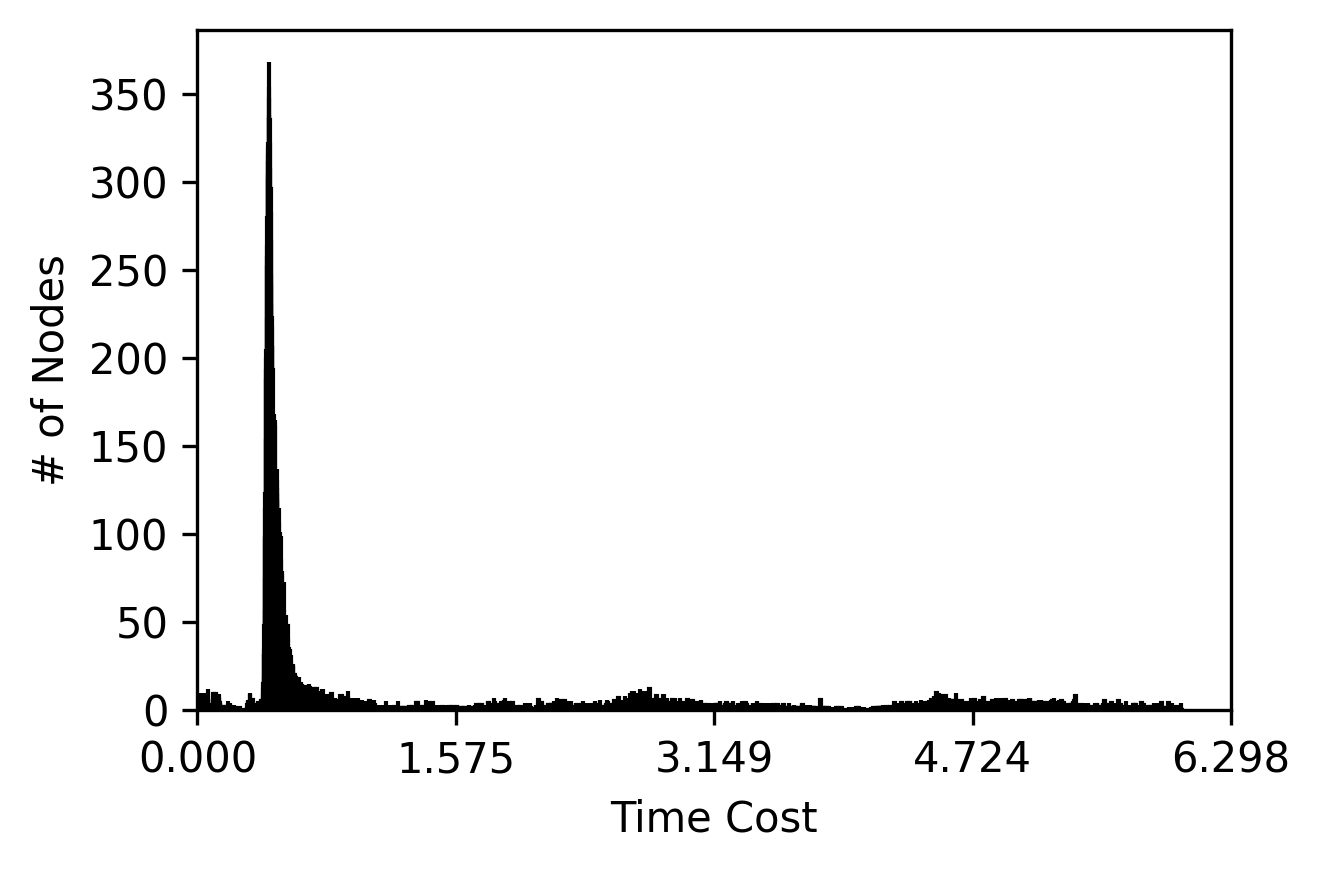}%
  \label{without poi overhead}}
  \caption{{\color{black}Distribution of anomaly scores and time costs of threat graphs sampled from various POI categories in E5-Clearscope.}}
\end{figure*}

\subsubsection{Threat Hunting Effectiveness} \label{Threat Hunting Effectiveness} 
{\color{black} 
We evaluate \textsc{ProHunter}'s adaptability across three categories of POIs, defined by their source and nature: 1) explicit IoCs \cite{megrapt}; 2) implicit suspicious nodes from anomalous alerts \cite{deephunter}; and 3) arbitrary system nodes \cite{provgsearcher,ghunter}. 
Since POI choice influences the semantic coverage of sampled threat graphs, each category is tested to assess its impact on hunting accuracy.

\noindent \textbf{Performance w/ Explicit IoCs.}
This category comprises CTI-documented IoCs, such as payload names or C\&C addresses.
As summarized in Table \ref{threat hunting with ioc}, \textsc{ProHunter} accurately matches APT attacks across 6 datasets and 28 attack cases with their corresponding CTI reports. Although a few benign subgraphs were occasionally sampled, their frequency was minimal and had negligible impact on the overall performance. Such superior performance indicates that IoCs are precise POIs, which can effectively locate attack subgraphs and filter benign interactions.
%Despite this success, the approach has a notable limitation: it remains vulnerable to circumvention by attackers who modify or evade IoCs. To mitigate this issue, \textsc{ProHunter} adopts a more flexible approach by relaxing the reliance on static IoCs.

\begin{comment}
\begin{figure}[t]
  \centering
  \includegraphics[scale=0.13]{poi_quality.png}
  \caption{The anomaly score distribution of sampled threat graphs on alarms in E5-Clearscope.}
  \label{poi quality}
\end{figure}

\begin{figure}[t]
  \centering
  \includegraphics[scale=0.13]{poi_random.png}
  \caption{The anomaly score distribution of sampled subgraphs w/o POI guidance in E5-Clearscope.}
  \label{without poi}
\end{figure}

\begin{figure}[t]
  \centering
  \includegraphics[scale=0.13]{poi_random_overhead.png}
  \caption{The time cost distribution of sampling threat graphs in E5-Clearscope.}
  \label{without poi overhead}
\end{figure}
\end{comment}

\noindent \textbf{Performance w/ Suspicious Nodes.}
%\textsc{ProHunter} can also utilize the propagation of suspicious information flows in the provenance graph to heuristically sample threat graphs, which allows it to process any type of POIs. 
In this experiment, anomalous nodes identified by our Recall-first pattern matching method (see \ref{poi}) are used as POIs. Note that anomalous nodes contain both true alarms (malicious POIs) and false alarms (benign POIs).
Threat graphs anchored on benign POIs are expected to be different from CTI-derived query graphs (i.e., lower anomaly scores), whereas graphs anchored on malicious POIs should capture attack behaviors (i.e., higher anomaly scores). 
Using the E5-Clearscope dataset, we sample subgraphs anchored on each POI and match them against CTI-derived query graphs.
Figure \ref{poi quality} depicts the anomaly score distribution for both benign and malicious subgraphs, where a distinct separation is evident. This distinction enables a clear decision threshold for accurate hunting.

\begin{table}[t]
  \scriptsize
  \centering
  \caption{{\color{black}\centering{Threat hunting based on explicit IoCs.}}}
  \label{threat hunting with ioc}
  \setlength{\tabcolsep}{3.2pt} % 减少列间距
  \begin{tabular}{|l|cccccc|}
    \hline
    {\color{black} \textbf {Scenario}} & \multicolumn{1}{c|}{{\color{black} \textbf{E3-Cadets}}} & \multicolumn{1}{c|}{{\color{black} \textbf{E3-Theia}}} & \multicolumn{1}{c|}{{\color{black} \textbf{E3-Trace}}} & \multicolumn{1}{c|}{{\color{black} \textbf{E5-Theia}}} & \multicolumn{1}{c|}{{\color{black} \textbf {E5-Clearscope}}} & {\color{black} \textbf{OpTC}} 
    \\  \hline  \multicolumn{1}{c}{} \vspace{-6.5pt} \\\hline
    {\color{black} Attack} & \multicolumn{1}{c|}{{\color{black} 4/4}} & \multicolumn{1}{c|}{{\color{black} 5/5}} & \multicolumn{1}{c|}{{\color{black} 2/2}} & \multicolumn{1}{c|}{{\color{black} 2/2}} & \multicolumn{1}{c|}{{\color{black} 2/2}} & {\color{black} 13/13} \\ \hline
    {\color{black} Benign} & \multicolumn{1}{c|}{{\color{black} 0/11}} & \multicolumn{1}{c|}{{\color{black} 0/17}} & \multicolumn{1}{c|}{{\color{black} 0/8}} & \multicolumn{1}{c|}{{\color{black} 0/6}} & \multicolumn{1}{c|}{{\color{black} 0/8}} & {\color{black} 0/35} \\ \hline
    \end{tabular}
\end{table}

\noindent \textbf{Performance w/o POIs.}
\textsc{ProHunter} supports exhaustive threat hunting by specifying arbitrary system entities as POIs (i.e., no indicators available). In this setting, both benign subgraphs and threat graphs are sampled. Figure \ref{without poi} presents the resulting distribution of subgraph anomaly scores, with benign subgraph anomaly scores clustering below 0, markedly lagging behind attack graph scores that are typically above 0.6. This pronounced gap indicates that \textsc{ProHunter} can effectively isolate malicious patterns even without prior knowledge.
Moreover, the sampling process is efficient. Figure~\ref{without poi overhead} presents that most nodes are processed within 1 second, with a maximum of ~6 seconds.
On average, processing 30,860 nodes requires only 0.4 seconds per graph, making the approach feasible for real-time or large-scale deployments.
%These capabilities enable automated detection and prioritization of high-risk entities with superior precision and efficiency, making \textsc{ProHunter} valuable for large-scale network environments.

These results reveal that, unlike existing methods reliant on explicit IoCs, \textsc{ProHunter} preserves strong threat hunting efficacy across diverse POI categories, while achieving fast, scalable performance suitable for large-scale environments.
}

{\color{black}
\subsection{Ablation Study} \label{abl}
%Semantic gap issues challenge the practicality of threat hunting solutions.
\textsc{ProHunter} addresses the semantic gap issue by introducing feature enhancement techniques and adaptive graph representation methods. To validate the efficacy of these techniques, we perform ablation experiments across diverse datasets.

\begin{figure}[t]
  \centering
  \subfloat[\scriptsize{{\color{black}Features of MEGR-APT.}}]{\includegraphics[width=1.74in]{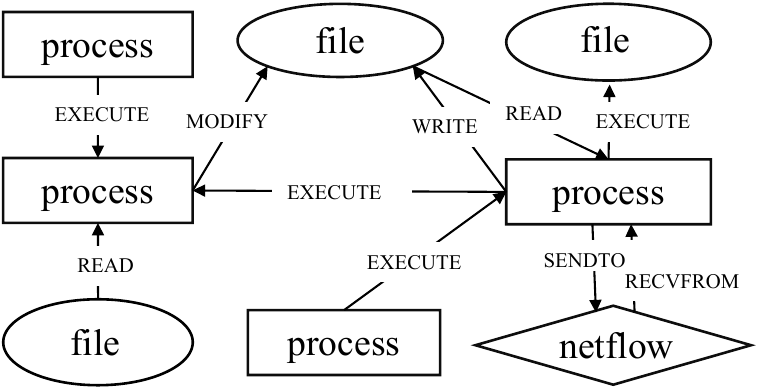}%
  \label{megrapt feature}}
  \hfil
  \subfloat[\scriptsize{{\color{black}Features of \textsc{ProHunter}.}}]{\includegraphics[width=1.74in]{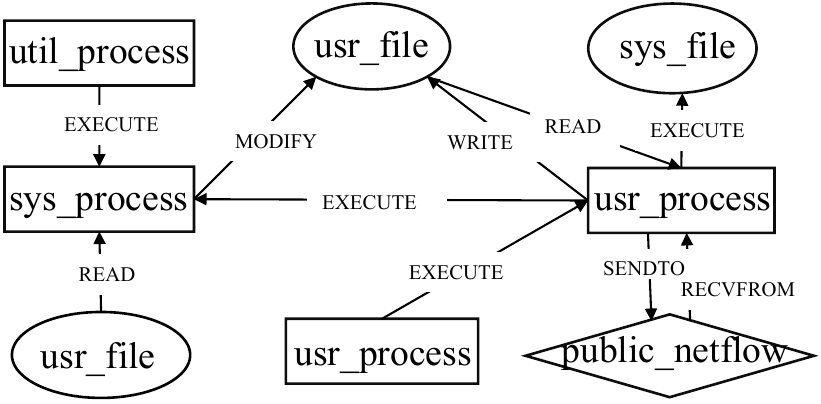}%
  \label{prohunter feature}}
  \caption{{\color{black}Feature initialization in a threat graph of E5-Theia.}}
\end{figure}

\subsubsection{Impact of Feature Enhancement}
We assess the role of abstracted node types in attack representation by replacing \textsc{ProHunter}'s feature initialization with MEGR-APT's coarse-grained scheme (i.e., process, file, netflow), keeping all other settings identical (e.g., GNN models, training and testing data). 
Table \ref{feature enhancement analysis} demonstrates the superior performance of \textsc{ProHunter}, particularly in FPR and accuracy. This improvement stems from the abstracted node features that capture deeper attack semantics and improve the representation discriminability. To illustrate, we visualized  threat graphs using both feature initialization methods in Figures \ref{megrapt feature} and \ref{prohunter feature}. The results indicate that MEGR-APT delivers shallower interaction semantics, making less discernible attack representations.
\begin{table}[t]
  \scriptsize
  \centering
  \caption{Effect of feature enhancement (w/ FE) and without feature enhancement (w/o FE) on threat hunting.}
  \label{feature enhancement analysis}
  \setlength{\tabcolsep}{5pt} % 减少列间距
  \begin{tabular}{|l|l|c|c|c|c|}
    \hline
    {\color{black} \textbf{Model}} & {\color{black} \textbf{Dataset}} & {\color{black} \textbf{Recall}}$\uparrow$ & {\color{black} \textbf{FPR}}$\downarrow$ & {\color{black} \textbf{Accuracy}}$\uparrow$ & {\color{black} \textbf{AUC}$\uparrow$}    
    \\  \hline  \multicolumn{1}{c}{} \vspace{-6.5pt} \\\hline
    {\color{black} } & {\color{black} E5-Theia} & {\color{black} 1.00} & {\color{black} 0.00} & {\color{black} 1.00} & {\color{black} 1.00} \\ \cline{2-6} 
    {\color{black} } & {\color{black} E5-Clearscope} & {\color{black} 1.00} & {\color{black} 0.00} & {\color{black} 1.00} & {\color{black} 1.00} \\ \cline{2-6} 
    \multirow{-3}{*}{{\color{black} w/ FE}} & {\color{black} OPTC} & {\color{black} 1.00} & {\color{black} 0.07} & {\color{black} 0.93} & {\color{black} 0.99} 
    \\  \hline  \multicolumn{1}{c}{} \vspace{-6.5pt} \\\hline
    {\color{black} } & {\color{black} E5-Theia} & {\color{black} 1.00} & {\color{black} 0.03} & {\color{black} 0.97} & {\color{black} 0.99} \\ \cline{2-6} 
    {\color{black} } & {\color{black} E5-Clearscope} & {\color{black} 1.00} & {\color{black} 0.04} & {\color{black} 0.96} & {\color{black} 0.99} \\ \cline{2-6} 
    \multirow{-3}{*}{{\color{black} w/o FE}} & {\color{black} OPTC} & {\color{black} 0.94} & {\color{black} 0.08} & {\color{black} 0.92} & {\color{black} 0.97} \\ \hline
    \end{tabular}
\end{table}

\subsubsection{Effect of Adaptive Graph Representation} \label{threshold flexibility}
Unlike prior static graph representation that captures only globalized attack semantic similarities, \textsc{ProHunter} employs adaptive graph representation to mine additionally localized similarities.
This experiment evaluates both approaches on datasets where over 99\% of graphs are from benign POIs, with threat graphs being a rare minority.
%Their threat thresholds $\theta$ were set at 0.3 and 0.45, corresponding to the points of highest Recall.

Table \ref{graph matching analysis} reveals that static graph representation leads to higher FPRs due to its limited ability to distinguish threat graphs from query graphs. In contrast, \textsc{ProHunter} reduces FPR by nearly 10\% on the OpTC dataset and achieves zero FPR across E3 datasets by extracting fine-grained similarities in attack patterns.
Furthermore, its higher Recall and accuracy highlight \textsc{ProHunter}'s ability to accurately identify threats and resist most false positives. This effectively mitigates semantic gaps between attack behaviors recorded in CTIs and those captured in audit logs.
\begin{table}[!t]
  \scriptsize
  \centering 
  \caption{Effect of adaptive and static graph representation methods on threat hunting.}
  \label{graph matching analysis}
  \setlength{\tabcolsep}{4pt} % 减少列间距
  \begin{tabular}{|c|l|c|c|c|c|}
  \hline
  \textbf{Model} & \textbf{Dataset} & \textbf{Recall}$\uparrow$ & \textbf{FPR}$\downarrow$ & \textbf{Accuracy}$\uparrow$ & \textbf{AUC}$\uparrow$ 
  \\  \hline  \multicolumn{1}{c}{} \vspace{-6.5pt} \\\hline
  \multirow{4}{*}{\begin{tabular}[c]{@{}c@{}}Adaptive Graph \\ Representation\end{tabular}} & E3-Cadets & 1.00 & 0.00 & 1.00 & 1.00 \\ \cline{2-6} 
   & E3-Theia & 1.00 & 0.00 & 1.00 & 1.00 \\ \cline{2-6} 
   & E3-Trace & 1.00 & 0.00 & 1.00 & 1.00 \\ \cline{2-6} 
   & OpTC & 1.00 & 0.07 & 0.93 & 0.99 
   \\  \hline  \multicolumn{1}{c}{} \vspace{-6.5pt} \\\hline
  \multirow{4}{*}{\begin{tabular}[c]{@{}c@{}}Static Graph \\ Representation\end{tabular}} & E3-Cadets & 1.00 & 0.00 & 1.00 & 1.00 \\ \cline{2-6} 
   & E3-Theia & 1.00 & 0.01 & 0.99 & 1.00 \\ \cline{2-6} 
   & E3-Trace & 1.00 & 0.03 & 0.97 & 0.99 \\ \cline{2-6} 
   & OpTC & 0.98 & 0.16 & 0.84 & 0.98 \\\hline
  \end{tabular}
  \end{table}

\begin{figure}[!t]
  \centering
  \color{black}
  \includegraphics[scale=0.15]{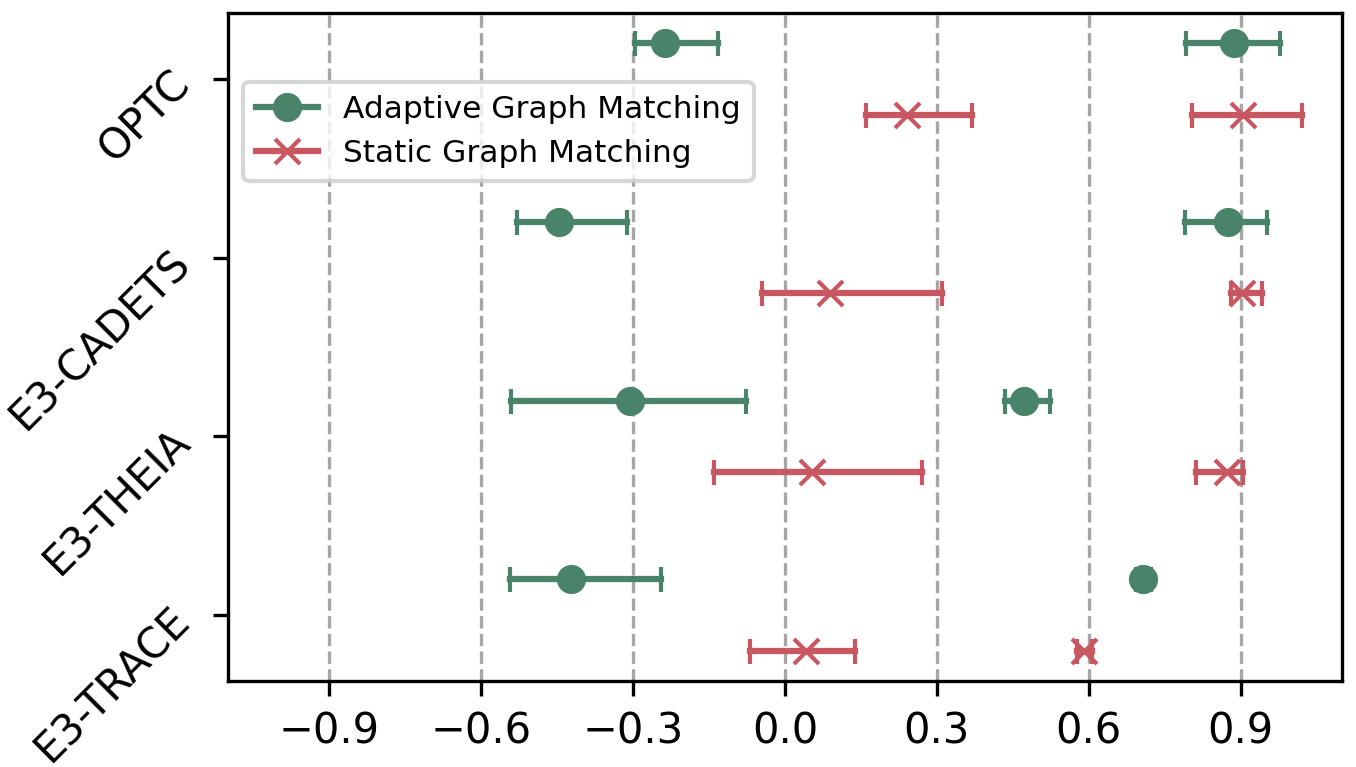}
  \caption{Threat scores of benign (left) and threat (right) graphs at the 10th, 50th, and 90th percentiles using adaptive and static graph representations.}
  \label{graph matching ablation figure}
\end{figure}

{\color{black}
\noindent \textbf{Flexible Setting of Threshold $\theta$.} We further investigate the sensitivity of these two methods to the threat thresholds. As illustrated in Figure \ref{graph matching ablation figure}, both methods assign higher threat scores to threat graphs. 
However, compared to the static representation method, \textsc{ProHunter} yields lower scores for benign graphs, as their weaker attack semantics are effectively identified by our adaptive representation mechanism.
This widening gap in threat scores between benign and threat graphs offers greater flexibility in adjusting the threshold $\theta$, facilitating more precise control over the balance between FPR and Recall with minimal trade-offs.
Specifically, setting $\theta$ of \textsc{ProHunter} between 0 and 0.3 ensures a Recall of 100\%, with only a modest increase in FPR to around 2\%-7\%. Conversely, adjusting $\theta$ to 0.5 lowers the maximum FPR to just 1\%, with a slight trade-off of missing one true threat in the E3-Theia. These results underscore the adaptability of \textsc{ProHunter} in effective threat hunting by adjusting $\theta$ across a wide range.
}
%These results demonstrate that \textsc{ProHunter} widens the threat score gap between malicious and benign threat graphs by extracting localized and globalized attack semantics, thereby enhancing its threat hunting effectiveness.
%This enhanced contrast simplifies threshold setting, significantly boosting the model's discriminative ability.
%These improvements underscore the effectiveness of \textsc{ProHunter} in extracting inherent attack semantics from threat graphs by comparing localized behavior patterns with query graphs.
% through intra- and inter-message passing mechanisms.

%The semantic gap arises from inconsistencies between CTI and audit logs in attack descriptions, potentially undermining the effectiveness of threat hunting. 
%\textsc{ProHunter} addresses this gap by first abstracting low-level entity descriptions to a unified level through a node abstraction strategy, then mining consistent localized attack semantics via an inter-message passing mechanism. 
%As shown in Table \ref{graph matching analysis}, \textsc{ProHunter} achieves a higher AUC score compared to previous globalized attack representations, indicating its ability to bridge the semantic gap.

\subsection{Comparative Experiment} \label{comp exp}
\subsubsection{Graph Compression Comparison}
We first compare the compaction performance of \textsc{ProHunter} against Sleuth~\cite{sleuth}.
Since Sleuth is not open-sourced, its memory usage is estimated from the descriptions in~\cite{sleuth}. Specifically, Sleuth encodes edges in variable length, averaging 10 bytes per edge for estimation. Node memory usage is also calculated following Sleuth's storage conventions, accounting for identifiers, names, and types. To ensure fairness, all datasets in Table~\ref{ppg analysis} are deduplicated before measurement.
The evaluation datasets listed in Table \ref{ppg analysis} have been deduplicated for fairness.

{\color{black}
The results in Table~\ref{comp ppg} show that \textsc{ProHunter} achieves 47\%-75\% higher compaction efficiency across all datasets. The gain is most pronounced in E5-Clearscope dataset, where uneven node degrees are efficiently handled by our hierarchical encoding scheme: only a small subset of dependency-explosion nodes is stored in extended structures, while sparse nodes and edges are kept in lightweight formats.
}
Meanwhile, node compaction via semantic abstraction significantly reduces memory cost, up to 60\% in datasets like E3-Trace with large-scale nodes.
In short, \textsc{ProHunter} optimizes memory usage, yielding compact yet information-rich provenance graphs.
{\color{black}
This comparison can be viewed as an ablation-style evaluation demonstrating the effectiveness of PPG's hierarchical edge compression and node-level semantic abstraction. To this end, we do not introduce an additional ablation study for PPG.
}

\begin{table}[t]
  \scriptsize
  \centering
  \caption{Comparison of \textsc{ProHunter} and Slueth in provenance graph compaction.}
  \label{comp ppg}
  \setlength{\tabcolsep}{4.2pt} % 减少列间距
  \begin{tabular}{|l|c|c|c|c|c|}
  \hline
  \multirow{2}{*}{\textbf{Dataset}} & \multicolumn{2}{c|}{\textbf{\begin{tabular}[c]{@{}c@{}}Total Mem.  (MB)\end{tabular}}} & \multicolumn{2}{c|}{\textbf{\begin{tabular}[c]{@{}c@{}}Daily Mem. (MB/Day)\end{tabular}}} & \multirow{2}{*}{\textbf{\begin{tabular}[c]{@{}c@{}}Reduction \\ (\%)\end{tabular}}} \\ \cline{2-5}
   & \multicolumn{1}{c|}{\textbf{Sleuth($\approx$)}} & \textbf{\textsc{ProHunter}} & \multicolumn{1}{c|}{\textbf{Sleuth($\approx$)}} & \textbf{\textsc{ProHunter}} &     
   \\  \hline  \multicolumn{1}{c}{} \vspace{-6.5pt} \\\hline
   E3-Cadets & \multicolumn{1}{c|}{36.83} & \textbf{17.27} & \multicolumn{1}{c|}{7.29} & \textbf{3.42} & 
  \textbf{53.09\%$\downarrow$} \\\hline
  E3-Theia & \multicolumn{1}{c|}{90.39} & \textbf{37.12} & \multicolumn{1}{c|}{28.51} & \textbf{11.71} & \textbf{58.92\%$\downarrow$}\\\hline
  E3-Trace & \multicolumn{1}{c|}{462.23} & \textbf{184.00} & \multicolumn{1}{c|}{43.40} & \textbf{17.28} & \textbf{60.20\%$\downarrow$ } \\\hline
  {\color{black}E5-Theia} & {\color{black}76.14} & {\color{black}\textbf{40.10}} & {\color{black}48.09} & {\color{black}\textbf{25.26}} & {\color{black}\textbf{47.47\%$\downarrow$}}
  \\ \hline
  {\color{black}E5-Clearscope1} &  {\color{black}22.85} & {\color{black}\textbf{6.21}} & {\color{black}12.75} & {\color{black}\textbf{3.47}} & {\color{black}\textbf{72.78\%$\downarrow$}} \\ \hline
  {\color{black}E5-Clearscope2} & {\color{black}46.24} & {\color{black}\textbf{11.25}} & {\color{black}12.17} & {\color{black}\textbf{2.97}} & {\color{black}\textbf{75.60\%$\downarrow$}} \\ \hline
  OpTC & \multicolumn{1}{c|}{159.67} & \textbf{67.05} & \multicolumn{1}{c|}{43.75} & \textbf{18.37} & \textbf{58.01\%$\downarrow$}   \\ \hline
  \end{tabular}
\end{table}
\begin{table}[!]
  \scriptsize
  \centering
  \caption{{\color{black} Impact of different threat graph sampling strategies on threat hunting performance. * denotes the use of threat graph sampled by MEGR-APT}}
  \label{effect of graph sampling}
  \setlength{\tabcolsep}{5pt} % 减少列间距
  \begin{tabular}{|l|l|c|c|c|c|}
    \hline
    {\color{black} \textbf{Dataset}} & {\color{black} \textbf{Model}} & {\color{black} \textbf{Recall} $\uparrow$} & {\color{black} \textbf{FPR}$\downarrow$} & {\color{black} \textbf{Accuracy}$\uparrow$} & {\color{black} \textbf{AUC}$\uparrow$}   \\  \hline  \multicolumn{1}{c}{} \vspace{-6.5pt} \\\hline
    {\color{black} } & {\color{black} \textbf{\textsc{ProHunter}}} & {\color{black} \textbf{1.00}} & {\color{black} \textbf{0.00}} & {\color{black} \textbf{1.00}} & {\color{black} \textbf{1.00}} \\ \cline{2-6} 
    {\color{black} } & {\color{black} \textsc{ProHunter}*} & {\color{black} 0.50} & {\color{black} 0.00} & {\color{black} 1.00} & {\color{black} 1.00} \\ \cline{2-6} 
    {\color{black} } & {\color{black} MEGR-APT} & {\color{black} 1.00} & {\color{black} 0.03} & {\color{black} 0.98} & {\color{black} 0.98} \\ \cline{2-6} 
    \multirow{-4}{*}{{\color{black} E5-Theia}} & {\color{black} MEGR-APT*} & {\color{black} 0.50} & {\color{black} 0.04} & {\color{black} 0.96} & {\color{black} 0.97}    
    \\  \hline  \multicolumn{1}{c}{} \vspace{-6.5pt} \\\hline
    {\color{black} } & {\color{black} \textbf{\textsc{ProHunter}}} & {\color{black} \textbf{1.00}} & {\color{black} \textbf{0.00}} & {\color{black} \textbf{1.00}} & {\color{black} \textbf{1.00}} \\ \cline{2-6} 
    {\color{black} } & {\color{black} \textsc{ProHunter}*} & {\color{black} 0.50} & {\color{black} 0.02} & {\color{black} 0.98} & {\color{black} 0.98} \\ \cline{2-6} 
    {\color{black} } & {\color{black} MEGR-APT} & {\color{black} 1.00} & {\color{black} 0.06} & {\color{black} 0.94} & {\color{black} 0.99} \\ \cline{2-6} 
    \multirow{-4}{*}{{\color{black} E5-Clearscope}} & {\color{black} MEGR-APT*} & {\color{black} 0.50} & {\color{black} 0.07} & {\color{black} 0.93} & {\color{black} 0.94} \\ \hline
    \end{tabular}
  \end{table}
\subsubsection{Threat Graph Sampling Comparison}
We next benchmark \textsc{ProHunter} against the following three threat graph sampling algorithms, using the metrics in Section \ref{sec:metric}:
\begin{itemize}
  \item \textbf{MEGR-APT} \cite{megrapt}: Samples a set of IoCs and their linked process nodes within hop $k$ to construct a threat graph. Furthermore, the sampled graph size is constrained to match the query graph, mitigating dependency explosion.
  \item \textbf{WATSON} \cite{watson}: Extracts behavior instances of a process using bidirectional depth-first search. Forward traversal continues until a dependency explosion node is visited, while backward traversal samples first-order neighbors.
  \item \textbf{DeepHunter} \cite{deephunter}: Uses BFS to sample anomalous nodes (e.g., IoCs or EDR alerts) and their linked process nodes until all anomalous nodes are included or a predefined hop limit is exceeded.
\end{itemize}

{\color{black}
The comprehensive comparative results in Figure \ref{sampling comparison} reveal that \textsc{ProHunter} achieves the highest coverage of anomalous interactions ($>$ 70\%), while keeping the lowest noise rate ($<$ 20\%). By explicitly tracking suspicious information flows between anomalous entities, it preserves the contextual attack semantics.
Competing methods, however, often struggle to balance attack semantic coverage and noise rate effectively.
For example, DeepHunter and MEGR-APT rely on a rich set of IoCs and linked process nodes to cover critical attack semantics, which often inflate subgraphs and raise noise.
}
%At the same time, the overall as well as local semantics are captured by using the graph-matching mechanism of \textsc{ProHunter} to ensuring the integrity of the contextual semantics of key entities, the introduction of some noisy nodes does not affect the attack subgraph determination results. The relevant experimental proofs are presented in Section B.3.

\begin{figure}[!t]
  \centering
  \subfloat[\small{E3-Trace}]{\includegraphics[width=1.62in]{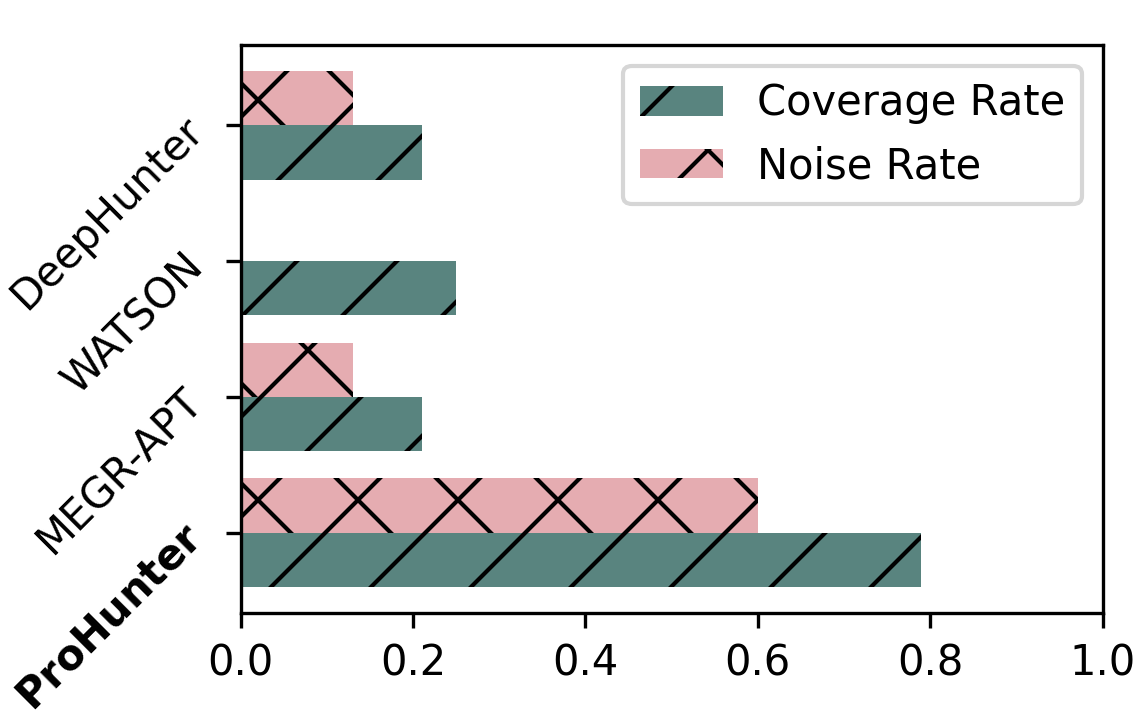}%
  \label{sampling comparison of trace}}
  \hfil
  \subfloat[\small{E3-Theia}]{\includegraphics[width=1.6in]{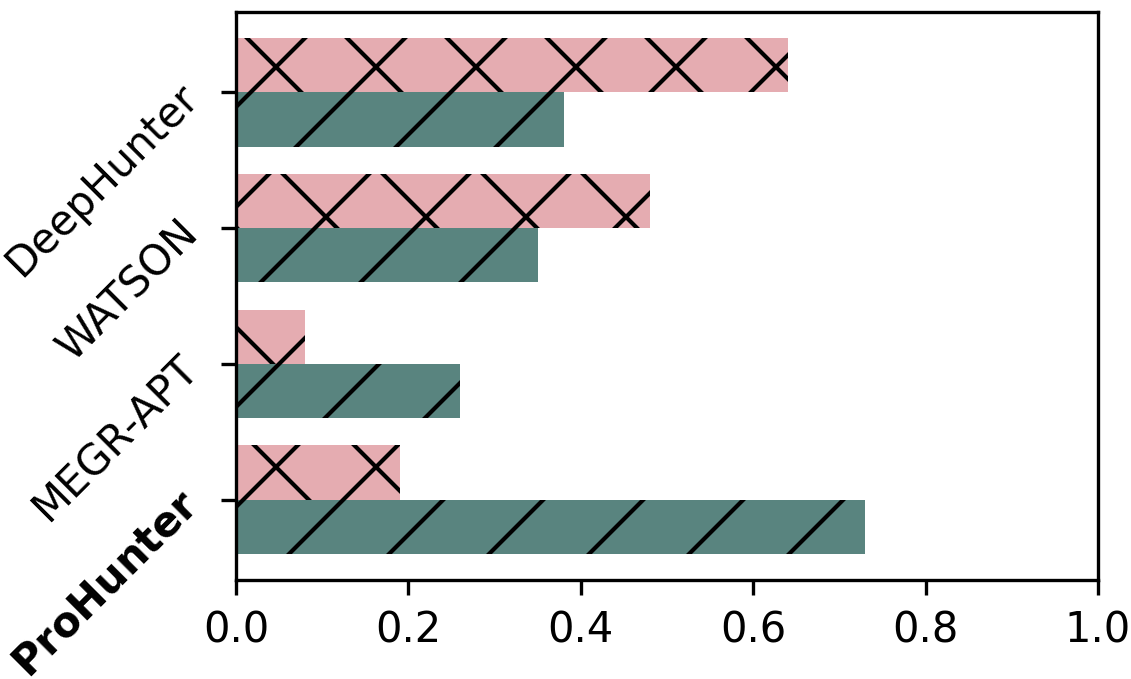}%
  \label{sampling comparison of theia}}
  \hfil
  \subfloat[\small{E3-Cadets}]{\includegraphics[width=1.6in]{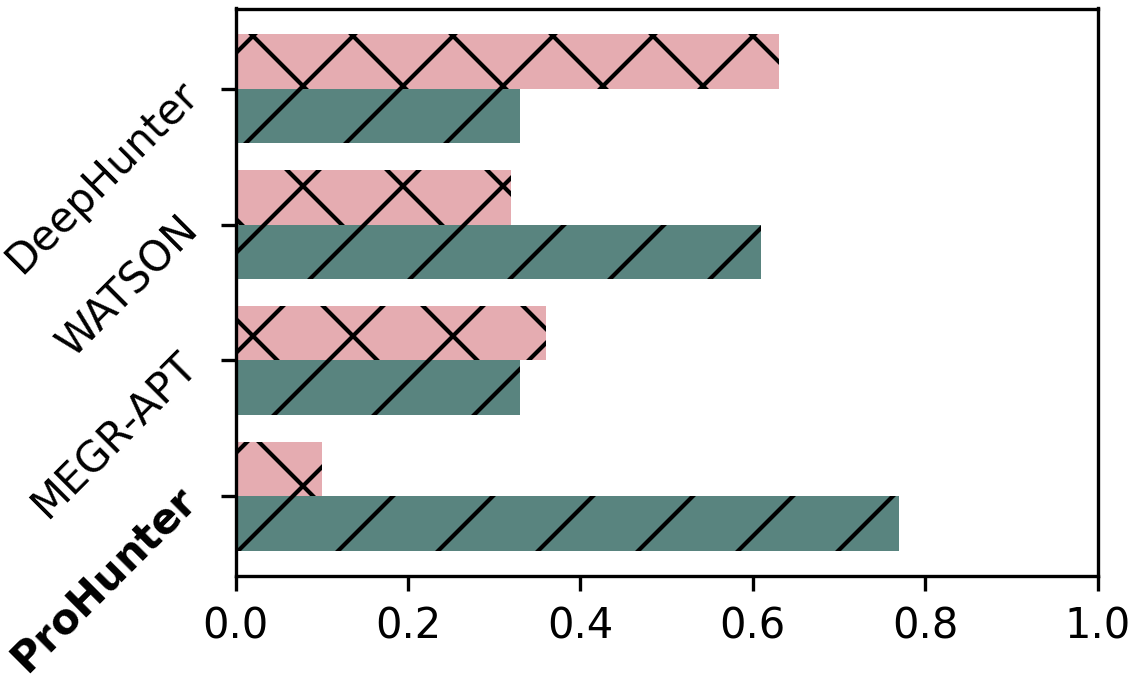}%
  \label{sampling comparison of cadets}}
  \hfil
  \subfloat[{\color{black}\small{E5-Theia}}]{\includegraphics[width=1.6in]{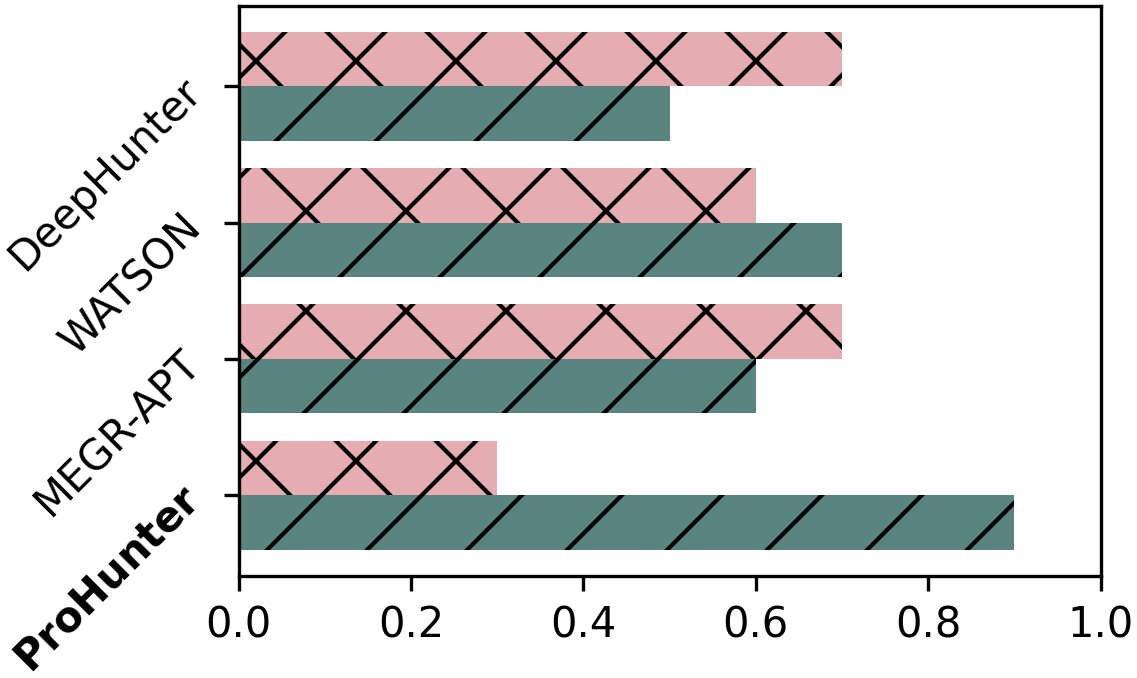}%
  \label{sampling comparison of cadets}}
  \hfil
  \subfloat[{\color{black}\small{E5-Clearscope}}]{\includegraphics[width=1.6in]{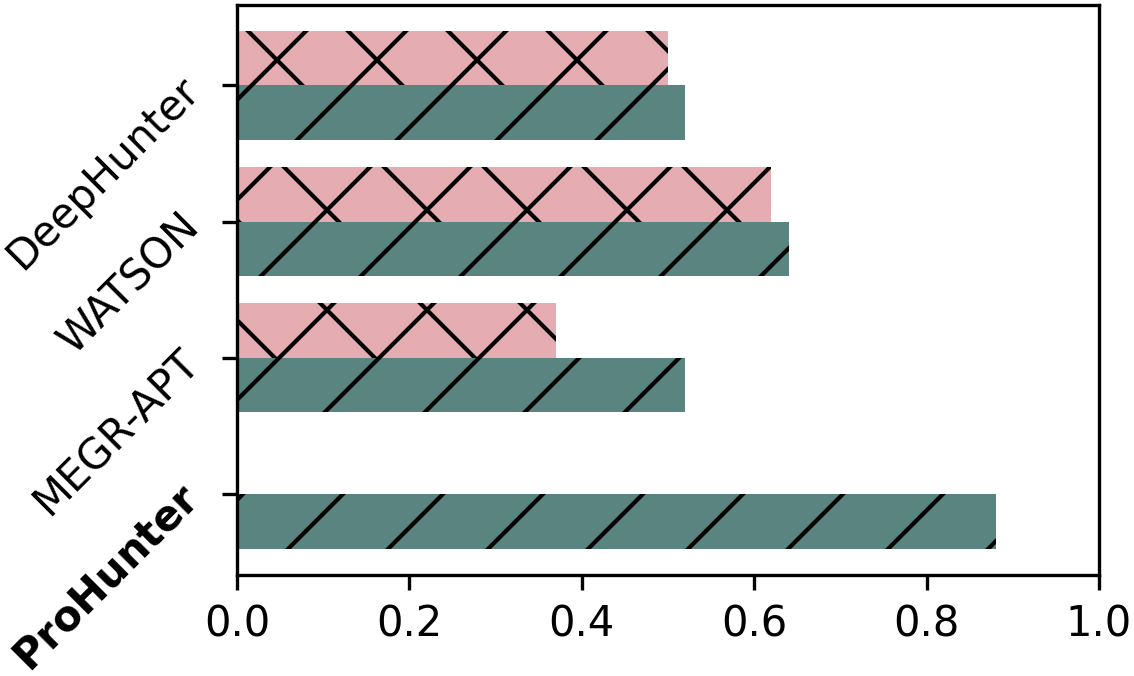}%
  \label{sampling comparison of cadets}}
  \hfil
  \subfloat[\small{OpTC}]{\includegraphics[width=1.6in]{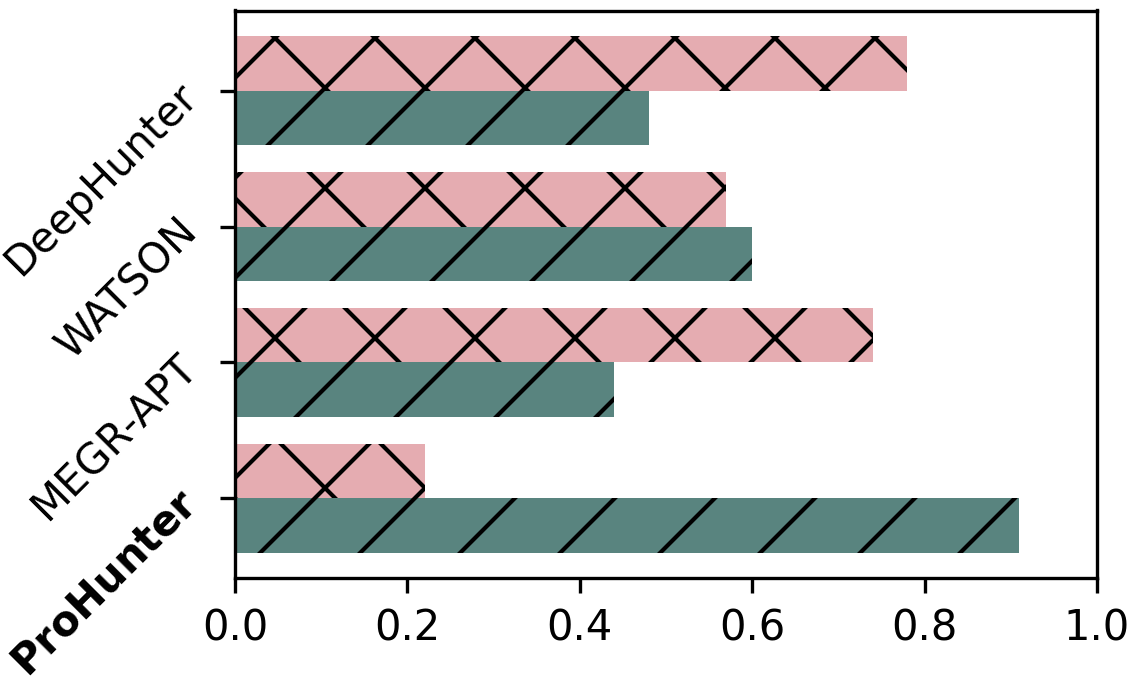}%
  \label{sampling comparison of optc}}
  \caption{Comparison of threat graph sampling algorithms.}
  \label{sampling comparison}
\end{figure}
\begin{figure*}[!h]
  \centering
  \subfloat[\small{{\color{black}Threat graph sampled by \textsc{ProHunter}.}}]{\includegraphics[width=3.6in]{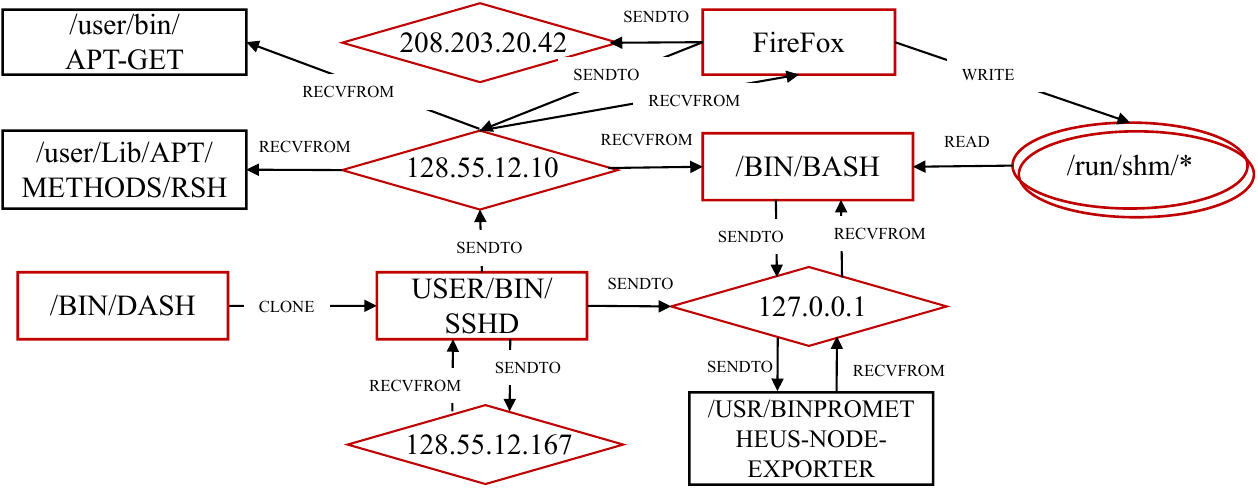}%
  \label{threat_graph_prohunter}}
  \hfil
  \subfloat[\small{{\color{black}Threat graph sampled by MEGR-APT.}}]{\includegraphics[width=3.4in]{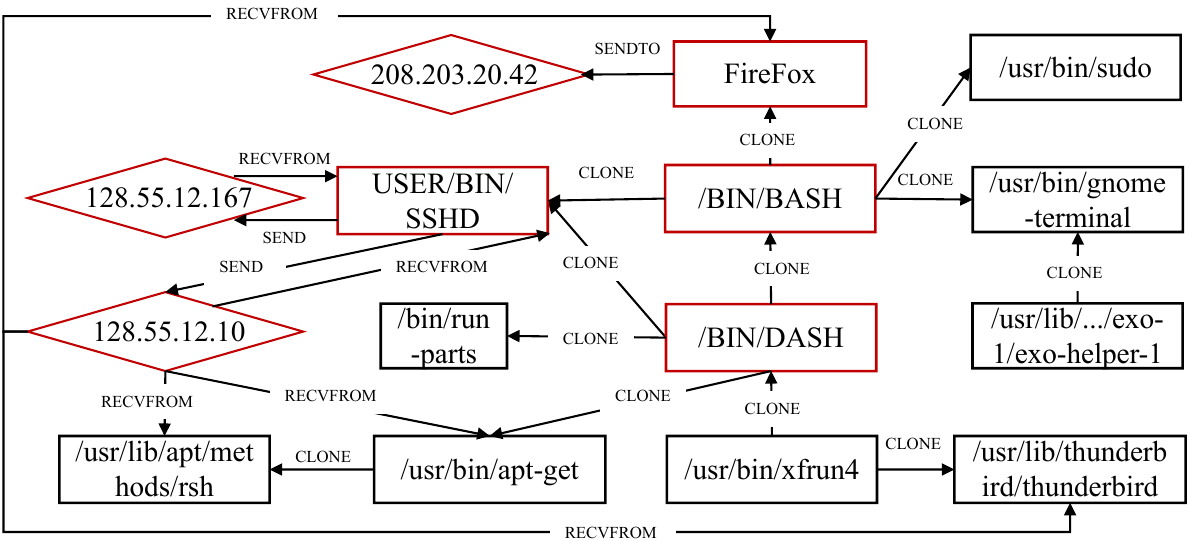}%
  \label{threat_graph_megr}}
  \caption{{\color{black}Threat graph comparison for E5-Theia. Red nodes indicate attack entities, while black nodes denote benign entities.}
  }
\end{figure*}

\noindent \textbf{Impact on Threat Hunting.}
We further assess the impact of sampling strategies on threat hunting using two E5 datasets. The experiments consider two threat hunting scenarios: threat graphs sampled by \textsc{ProHunter} and those sampled by MEGR-APT. The results of the effects of the two strategies on their respective performance are summarized in Table \ref{effect of graph sampling}.

The performance gains of MEGR-APT and \textsc{ProHunter} highlight the effectiveness of our sampling heuristics in enhancing threat hunting performance, as evidenced by higher Recall and fewer false positives.
Visual inspection (Figures~\ref{threat_graph_prohunter} and \ref{threat_graph_megr}) confirms the reason: our sampled threat graphs introduce only 2–3 extra nodes or edges while retaining complete attack semantics. MEGR-APT, in contrast, misses certain critical behaviors, such as file operations, due to rigid size limits and IoC over-dependence. Overall, our threat graph sampling algorithm not only enhances the threat hunting but also provides security analysts with more actionable insights.
{\color{black}
This experiment serves as an ablation study to evaluate the effectiveness of our heuristic threat graph sampling algorithm in improving downstream threat hunting performance. Therefore, we do not add a separate ablation section.
}

{\color{black}
\subsubsection{Threat Hunting Comparison} 
In this experiment, \textsc{ProHunter} is compared against three standard GNN variants and two state-of-the-art threat hunting systems across DAPRA datasets. The baselines include:
\begin{itemize}
  \item \textbf{GNN Variants}: Graph Convolutional Network (GCN) \cite{gcn}, GraphSAGE \cite{graphsage}, and Graph Attention Network (GAT) \cite{gat} which serve as generic graph representation learners. These models lack the adaptive graph representation mechanisms, making them suitable baselines for evaluating the effectiveness of our domain-specific designs.
  \item \textbf{MEGR-APT} \cite{megrapt}: Uses a Relational Graph Convolutional Network (RGCN) to obtain graph embeddings through normalized aggregation of neighboring node types, edge types and directions, followed by graph matching prediction via a Neural Tensor Network (NTN).
  \item \textbf{ProvG-Searcher} \cite{provgsearcher}: Treats threat hunting as a subgraph entailment problem, achieving efficient neural subgraph matching by combining GNN and order embedding techniques. It predicts subgraph relationships between process-centered provenance subgraphs and query graphs.
\end{itemize}

Table~\ref{threat hunting comp} reports the results. Across all DARPA datasets, \textsc{ProHunter} detects all attacks while maintaining the lowest FPRs. In contrast, the standard GNN baselines (GCN, GraphSAGE, and GAT) exhibit high FPRs and unstable recall. This limitation stems from their inability to capture the critical attack semantics inherent in threat graphs, making them susceptible to spurious matches within structurally similar benign subgraphs. 
Among the specialized threat hunting systems, ProvG-Searcher performs well on E3 datasets but fails to detect one APT campaign in E5-Clearscope, and its high FPR stems from mislabeling large benign graphs that partially resemble query patterns.
Our analysis shows that these benign graphs tend to be much larger than CTI-derived query graphs, increasing the probability of accidental structural alignment. 
%A potential solution to reduce false matches is to initialize finer-grained graph features to produce more distinctive representations.
MEGR-APT also exhibits elevated FPRs on E5-Theia and OpTC because its node initialization is based solely on coarse node types, which provides insufficient discrimination among semantically similar benign activities.
By contrast, \textsc{ProHunter} integrates semantic abstraction at the node level with an adaptive graph representation model that performs both intra- and inter-graph reasoning. This design enables it to capture intrinsic attack semantics, suppress superficial structural similarity, and robustly distinguish true threats from benign background behaviors.}
\begin{table}[t]
  \scriptsize
  \centering
  \caption{Comparison of threat hunting models.}
  \setlength{\tabcolsep}{5pt} % 减少列间距
  \color{black}
  \begin{tabular}{|c|c|c|c|c|c|}
  \hline
  \textbf{Dataset} & \textbf{Model} & \textbf{Recall} $\uparrow$& \textbf{FPR} $\downarrow$& \textbf{Accuracy} $\uparrow$& \textbf{AUC} $\uparrow$ 
  \\  \hline  \multicolumn{1}{c}{} \vspace{-6.5pt} \\\hline
  \multirow{6}{*}{E3-Cadets} & GAT & 0.67 & 0.11 & 0.89 & 0.97 \\ \cline{2-6} 
   & GraphSAGE & 1.00 & 0.18 & 0.82 & 0.98 \\ \cline{2-6} 
   & GCN & 0.67 & 0.19 & 0.81 & 0.94 \\ \cline{2-6} 
   & MEGR-APT & 1.00 & 0.00 & 1.00 & 1.00 \\ \cline{2-6} 
   & ProvG-Searcher & 1.00 & 0.06 & 0.94 & 0.98 \\ \cline{2-6} 
   & \textbf{\textsc{ProHunter}} & \textbf{1.00} & \textbf{0.00} & \textbf{1.00} & \textbf{1.00} 
   \\  \hline  \multicolumn{1}{c}{} \vspace{-6.5pt} \\\hline
  \multirow{6}{*}{E3-Theia} & GAT & 0.80 & 0.13 & 0.87 & 0.95 \\ \cline{2-6} 
   & GraphSAGE & 0.80 & 0.08 & 0.92 & 0.96 \\ \cline{2-6} 
   & GCN & 1.00 & 0.18 & 0.82 & 0.94 \\ \cline{2-6} 
   & MEGR-APT & 1.00 & 0.05 & 0.95 & 0.99 \\ \cline{2-6} 
   & ProvG-Searcher & 1.00 & 0.12 & 0.88 & 0.90 \\ \cline{2-6} 
   & \textbf{\textsc{ProHunter}} & \textbf{1.00} & \textbf{0.00} & \textbf{1.00} & \textbf{1.00} 
   \\  \hline  \multicolumn{1}{c}{} \vspace{-6.5pt} \\\hline
  \multirow{6}{*}{E3-Trace} & GAT & 0.50 & 0.08 & 0.92 & 0.92 \\ \cline{2-6} 
   & GraphSAGE & 1.00 & 0.03 & 0.97 & 0.98 \\ \cline{2-6} 
   & GCN & 1.00 & 0.11 & 0.89 & 0.96 \\ \cline{2-6} 
   & MEGR-APT & 1.00 & 0.00 & 1.00 & 1.00 \\ \cline{2-6} 
   & ProvG-Searcher & 1.00 & 0.09 & 0.91 & 0.95 \\ \cline{2-6} 
   & \textbf{\textsc{ProHunter}} & \textbf{1.00} & \textbf{0.00} & \textbf{1.00} & \textbf{1.00} 
   \\  \hline  \multicolumn{1}{c}{} \vspace{-6.5pt} \\\hline
  \multirow{6}{*}{E5-Theia} & GAT & 0.50 & 0.09 & 0.91 & 0.97 \\ \cline{2-6} 
   & GraphSAGE & 1.00 & 0.11 & 0.89 & 0.97 \\ \cline{2-6} 
   & GCN & 0.50 & 0.07 & 0.93 & 0.95 \\ \cline{2-6} 
   & MEGR-APT & 1.00 & 0.03 & 0.98 & 0.98 \\ \cline{2-6} 
   & ProvG-Searcher & 1.00 & 0.10 & 0.90 & 0.93 \\ \cline{2-6} 
   & \textbf{\textsc{ProHunter}} & \textbf{1.00} & \textbf{0.00} & \textbf{1.00} & \textbf{1.00} 
   \\  \hline  \multicolumn{1}{c}{} \vspace{-6.5pt} \\\hline
  \multirow{6}{*}{E5-Clearscope} & GAT & 0.50 & 0.17 & 0.83 & 0.90 \\ \cline{2-6} 
   & GraphSAGE & 1.00 & 0.14 & 0.86 & 0.92 \\ \cline{2-6} 
   & GCN & 0.50 & 0.15 & 0.85 & 0.90 \\ \cline{2-6} 
   & MEGR-APT & 1.00 & 0.06 & 0.94 & 0.99 \\ \cline{2-6} 
   & ProvG-Searcher & 0.50 & 0.15 & 0.85 & 0.91 \\ \cline{2-6} 
   & \textbf{\textsc{ProHunter}} & \textbf{1.00} & \textbf{0.00} & \textbf{1.00} & \textbf{1.00} 
   \\  \hline  \multicolumn{1}{c}{} \vspace{-6.5pt} \\\hline
  \multirow{6}{*}{OPTC} & GAT & 1.00 & 0.60 & 0.41 & 0.83 \\ \cline{2-6} 
   & GraphSAGE & 0.87 & 0.44 & 0.56 & 0.84 \\ \cline{2-6} 
   & GCN & 0.87 & 0.48 & 0.53 & 0.88 \\ \cline{2-6} 
   & MEGR-APT & 0.93 & 0.08 & 0.92 & 0.96 \\ \cline{2-6} 
   & ProvG-Searcher & 0.87 & 0.17 & 0.83 & 0.89 \\ \cline{2-6} 
   & \textbf{\textsc{ProHunter}} & \textbf{1.00} & \textbf{0.07} & \textbf{0.93} & \textbf{0.99} \\ \hline
  \end{tabular}
  \label{threat hunting comp}
\end{table}

\subsection{Overhead Analysis} \label{sec:overhead}
We profile \textsc{ProHunter}'s runtime and memory footprint, focusing on PPG construction, threat graph sampling, and attack representation. The reported results are averaged over repeated processing of 5-minute event windows, each followed by a single hunting operation.
% representing the upper bound of performance overhead for the DARPA TC dataset.

%DARPA TC E3-Trace dataset as an example, which contains largest graph sizes within all datasets, 
%The results are averaged from overhead of PPG processing 5-minute windows of event batches at a time and performing threat hunting on one threat graph at a time.

%The training time of the adaptive graph representation and matching component has quadratic complexity, but it is a one-time effort. Since it operates on static graph structures with fixed node and edge attributes, it remains unaffected by \textit{concept drift} \cite{concept_drift}. 
Table~\ref{overhead} demonstrates that each 5-minute window ($\sim$4,200 events) is processed in 0.67 s on average, with the slowest case ($\sim$12,000 events) completing in under 5.4 s—less than 2\% of the wall-clock window time. Given a POI, threat graph sampling completes in 0.03 s and hunting in 0.18 s, achieving a throughput of $\sim$300 hunts/minute.
Memory usage remains modest: PPG storage requires only 0.06 MB per 5-minute window, with growth stabilizing over time. The attack representation model maintains a constant 556 MB footprint due to its dependencies and internal data structures. Overall, \textsc{ProHunter} executes hunting pipelines within seconds while incurring negligible memory overhead, making it highly scalable for real-time deployments.

\begin{table}[t]
  \scriptsize
  \centering
  \caption{System overhead of \textsc{ProHunter} on E3-Trace.}
  \label{overhead}
  \setlength{\tabcolsep}{3.5pt} % 减少列间距
  \begin{tabular}{|l|c|cc|}
  \hline
  \multirow{2}{*}{\textbf{\vspace{-0.3cm} Overhead}} & \multirow{2}{*}{\textbf{\vspace{-0.3cm} \begin{tabular}[c]{@{}c@{}}PPG Construction\\ (/ 5-minute window)\end{tabular}}} & \multicolumn{2}{c|}{\textbf{Threat Hunting (/ graph)}} \\ \cline{3-4} 
    &  & \multicolumn{1}{c|}{\textbf{\begin{tabular}[c]{@{}c@{}}Threat Graph \\ Sampling\end{tabular}}}& \textbf{\begin{tabular}[c]{@{}c@{}}Attack Representation \\ and Matching\end{tabular}}
    \\  \hline  \multicolumn{1}{c}{} \vspace{-6.5pt} \\\hline

  Time (s) & 0.67 & \multicolumn{1}{c|}{0.03} & 0.15 \\ \hline
  Memory (MB) & 0.06 & \multicolumn{1}{c|}{/} & 556.82 \\ \hline
  \end{tabular}
\end{table}

\subsection{Discussion}
\subsubsection{Requirements Fulfillment} 
The design of \textsc{ProHunter} addresses the practical needs outlined in Section \ref{introduction}, and experiments confirm consistent performance across diverse platforms.
For timeliness, provenance graphs are compacted to 3–25 MB/day, enabling real-time analysis and long-term APT lifecycle modeling with minimal overhead.
For adaptability and accuracy, the proposed sampling algorithm pinpoints APT activities hidden in extensive provenance graphs by tracing contextually suspicious information flows, then bridges semantic gaps between threat and query graphs through node abstractions and both localized and globalized attack representations. Furthermore, high throughput with minimal memory usage ensures scalable threat hunting.

%\textsc{ProHunter}'s memory-efficient PPG structure enables the storage of provenance graphs in main memory, ensuring subsequent real-time analysis tasks. The PPG structure exhibits effective compression performance across four cross-platform datasets, requiring less than 20 MB of memory per day to store provenance graphs, demonstrating its scalability. Additionally, PPG provides a unified framework for efficiently storing provenance graphs and includes reserved bits in its node and edge structures, facilitating future integration with additional provenance-based techniques  \cite{aptshield,conan}.
{\color{black}
\subsubsection{POI Selection}
We adopt a more general POI assumption that covers: 1) explicit IoCs; 2) implicit anomalous alarms; and 3) arbitrary system entities (no prior indicators). Unlike prior methods \cite{megrapt,poirot} that depend on explicit IoCs, \textsc{ProHunter} is independent of the specific POI category. As validated in Section \ref{Threat Hunting Effectiveness}, \textsc{ProHunter} maintains stable accuracy across all POI types, demonstrating its robustness. While compatible with any POI selection strategy, prioritizing high-anomaly POIs can further improve efficiency.
}

\subsubsection{Limitations} 
\textsc{ProHunter} depends on the integrity of audit logs and CTI reports. However, existing kernel event trackers~\cite{kellect,ebpfsp} still struggle to balance semantic completeness and efficiency \cite{kellect,ebpfsp}.
For example, OpTC dataset records shell commands but omits explicit entity dependencies, causing semantic loss.
Furthermore, CTI quality also impacts prior knowledge coverage and the effectiveness of threat hunting. While \textsc{ProHunter} mitigates these gaps, advances in audit logging and richer CTI reports will further improve threat hunting accuracy.

\section{Related Work}

{\color{black}
\subsection{Threat Hunting}
Threat hunting is a proactive defense practice where analysts iteratively generate and validate hypotheses to uncover APT activities that evade traditional detection systems~\cite{thtypes}. 
Some work automates the hypothesis generation process. For instance, AUTOMA~\cite{nour2024automa} performs knowledge discovery to derive attack hypotheses and their variants by mining correlations between CTI and telemetry data. Yi et al.~\cite{yi2024hypothesis} develop a five-step pipeline that leverages indicators of attack and IT asset information to construct and validate threat hypotheses. However, as CTIs and logs operate at different semantic levels, these methods often require manual interpretation or handcrafted rules to bridge the semantic gap.

A parallel line of work leverages whole-system provenance, which provides fine-grained causal dependencies between system entities, and combines it with CTI knowledge to operationalize hunting. Provenance-based approaches typically fall into two categories: prediction-based subgraph entailment and similarity-based graph matching.
Subgraph entailment prediction methods \cite{provgsearcher,ghunter,ma2025actminer} search for CTI-derived query graphs within the provenance graph. For example, ACTMINER~\cite{ma2025actminer} performs heuristic search with equivalent semantic transfer and causal relationship filtering. However, their accuracy is sensitive to graph size, leading to false positives or missed detections. 
Graph matching methods ~\cite{megrapt,deephunter,poirot} embed both sampled threat graphs and CTI-derived graphs for semantic similarity comparison. For instance, MEGR-APT~\cite{megrapt} learns attack representations to search for suspicious subgraphs in memory, issuing alerts based on semantic similarities.
Although more resilient to behavioral variations, they depend on IoC-driven subgraph sampling, making them vulnerable to mutation attacks.

Recent work explores cross-modal learning paradigms to directly bridge the semantic gap. CLIProv~\cite{li2025cliprov} aligns log semantics with threat intelligence via multimodal contrastive learning, and APT-CGLP~\cite{qiu2025apt} similarly aligns attack narratives with provenance graphs through graph–language contrastive training, thereby reducing the burden of manual CTI parsing. Other efforts operationalize CTIs more directly, such as AIThreatTrack~\cite{purba2025towards}, which converts CTI descriptions into executable Kibana queries using large language models to search attack traces in logs. Complementarily, TREC~\cite{lv2024trec} utilizes CTI reports to support fine-grained APT tactic/technique recognition from provenance graphs under a few-shot setting. 
While these methods improve knowledge utilization, they typically require substantial training data or complex data-generation pipelines.
}

\subsection{APT Detection}
Provenance-based APT detection aims to identify abnormal system activities, including zero-day attacks, by either leveraging expert knowledge to customize detection rules or employing deep learning models.
Existing detection methods operate at different granularities: graph-level and node-level. Graph-level methods~\cite{unicorn,provgem} detect anomalies through global structural and attribute changes. For example, UNICORN~\cite{unicorn} uses graph sketching to create fixed-size representations of streaming provenance graphs and employs evolutionary clustering to model normal system behavior changes, detecting anomalies as deviations from learned patterns. These methods enable rapid detection but lack fine-grained forensic insights into specific malicious entities. 
Node-level approaches~\cite{aptkgl,magic,threatrace,qiu2025provenance} analyze detailed contextual interactions for more accurate threat localization.
For instance, TRAP~\cite{qiu2025provenance} employs multi-scale graph representation learning with feature reconstruction and contrastive learning to capture behavioral semantics at different granularities, enabling better distinction between sophisticated benign activities and APT attack patterns. While achieving higher accuracy, these methods incur increased computational overhead due to fine-grained analysis.

Recent methods employ multi-level detection pipelines to balance efficiency and precision. For instance, MAGIC~\cite{magic} first isolates potentially anomalous subgraphs, then applies node-level detection for refined analysis. Similarly, Kairos~\cite{kairos} initially flags suspicious edges, then aggregates them into attack graphs for anomaly confirmation.
While these systems effectively identify zero-day attacks, they often suffer from high false positive rates and limited interpretability.
%In contrast, APT hunting systems proactively search for APT activities using prior attack knowledge from CTIs, enhancing both interpretability and accuracy.

\section{Conclusion}
{\color{black}
We present \textsc{ProHunter}, a comprehensive provenance-based threat hunting system. Our design couples a memory-efficient provenance graph representation capable of compressing daily audit logs to as little as 3–25 MB, with a generalized threat graph sampling algorithm that retains core malicious interactions while filtering benign noise.
To bridge the semantic gap between CTI-derived query graphs and audit log-derived provenance graphs, \textsc{ProHunter} leverages a tailored GIN-based attack representation model that integrates feature enhancement with intra- and inter-graph message passing, enabling robust threat hunting even in the absence of explicit IoCs.
Extensive evaluations on multiple DARPA datasets demonstrate that \textsc{ProHunter} outperforms leading threat hunting systems in accuracy and efficiency, proving its practicality for real-world deployment.
}

\section*{Acknowledgements}
This work has been partially supported by the National Natural Science Foundation of China (Grant Nos. 62372410, U22B2028), the Zhejiang Provincial Natural Science Foundation of China (Grant No. LZ23F020011), and the Zhejiang Province Leading Goose Program (Grant No. 2025C01013).


\bibliographystyle{elsarticle-num} 
\bibliography{ref}





% \end{thebibliography}

\appendix
\section{General Compression Strategies}\label{generic_compression}
%There are some semantic redundant events in audit logs, which can impact the efficiency of PPG construction. Therefore, 
We employ two general compression strategies to eliminate redundant events in batch audit logs: 1) removing consecutive duplicate events (denoted as $S1$) \cite{gsss}; and 2) filtering identical network events within fixed time windows (denoted as $S2$) \cite{versioning}. These strategies improve the efficiency of the PPG construction. 
$S1$ retains only the first occurrence of a single or pairs of repeated events when they occur consecutively. For example, if a process reads a file in chunks and writes to another file, creating consecutive duplicate `read' and `write' events in audit logs, only the initial `read', `write' and `read-write' event pair are kept (a similar scenario applies to network streams). $S2$ leverages a time window $T$ to retain network events at a low frequency. Specifically, for the same remote socket, only the first sending/receiving event within the time window $T$ is retained, while subsequent events are discarded. We set the value of $T$ to 5 minutes, corresponding to the time window in which the PPG receives batch events.

% \usepackage{multirow}
We applied the above compression strategies on DARPA datasets to evaluate its effectiveness and used these deduplicated datasets for PPG construction. The evaluation results are shown in Table \ref{deduplication_results}, which illustrates that $S1$ and $S2$ exhibit varying degrees of compression across different datasets. $S1$ significantly compresses E3 datasets, achieving compression ratios of 60\% to 90\%. 
This indicates that the audit tracking tools in E3 generate a lot of consecutive duplicate events.
% without employing reduction strategies  \cite{kellect}. 
%The compression of consecutive duplicate events in OpTC is less than 2\%, which may be due to audit frameworks differences and preliminary event reduction strategies implemented by OpTC. 
Second, the compression ratio of $S2$ in OpTC is 50\% to 60\%, reflecting a significant reduction in network events. This can be attributed to the large-scale network environment of the OpTC. After compression, the ratio of nodes and edges in each dataset is of the same order of magnitude.

\section{Miscellaneous Encoding Bits in PPG}\label{misc}
There are miscellaneous bits shown in Figure \ref{ppg} used for encoding provenance information and feature extensions. First, each edge contains 4-bit `type' to encode the edge type (see Table \ref{nodes edges}), and 1-bit `dir' to encode the edge direction. To ensure chronological traversal, each edge includes a 27-bit `timestamp' for encoding millisecond timestamp within a day and a 5-bit `date' for encoding relative day.
\footnote{This setting supports provenance graphs spanning up to a month,\\ with additional bits allowing for longer periods.} 
%\footnote{This setting supports adding bits to extend the storage period.} 
Second, we demonstrate how to integrate the version control mechanism  \cite{versioning,versioning_old} into our PPG component as an extended function. This mechanism can be leveraged to reduce semantically redundant events.
Specifically, the edge and node structures need to be assigned an additional 8-bit `version' to record the version number.
If the operated object's version remains unchanged from the last identical event, the current event will be discarded. Otherwise, the edge will be added to the queue. If it is an incoming edge (e.g., `write'), it indicates that the semantics of the object have changed and the object's version number will be incremented by 1. 
It is worth noting that in extended structures, the `version' attribute should be expanded to 16 bits to accommodate more complex dependencies. More functions can be incorporated by means of extending reserved bits \cite{aptshield}.

\section{Threat Graph Sampling Algorithm} \label{app:adaptive bfs}
Algorithm \ref{adaptive bfs} presents the pseudo code for our threat graph sampling algorithm. The algorithm accepts POI nodes and a predefined traversal depth $k$ as inputs, then performs BFS traversal on the PPG. The core sampling process (lines 2-21) extracts threat graphs centered on each POI through a customized BFS approach. Line 6 filters suspicious dependencies of visited nodes based on the sampling rules in Table \ref{sampling rules}, while lines 7-18 sample suspicious neighbors and recursively process other discovered POI nodes. Finally, the algorithm merges threat graphs with overlapping POIs (line 22) and consolidates duplicate named nodes within each graph (line 23).
\begin{table}[t]
  \scriptsize
  \centering
  \caption{\centering{The deduplication results of DARPA TC datasets.}}
  \label{deduplication_results}
  \begin{tabular}{|c|c|c|cc|c|}
    \hline
    \multirow{2}{*}{\textbf{Dataset}} &
      \multirow{2}{*}{\textbf{\#Entities}} &
      \multirow{2}{*}{\textbf{\#Events}} &
      \multicolumn{2}{c|}{\textbf{\# Events Reduction}} &
      \multirow{2}{*}{\textbf{\begin{tabular}[c]{@{}c@{}}\# Events\\ Remaining\end{tabular}}} \\ \cline{4-5}
     &
       &
       &
      \multicolumn{1}{c|}{\textbf{\begin{tabular}[c]{@{}c@{}}Reduced \\ By S1\end{tabular}}} &
      \textbf{\begin{tabular}[c]{@{}c@{}}Reduced \\ By S2\end{tabular}} &
      \\  \hline  \multicolumn{1}{c}{} \vspace{-6.5pt} \\\hline

    E3-Cadets & 259K & 7.4M   & \multicolumn{1}{c|}{5.1M}   & 6K   & 2.3M  \\ \hline
    E3-Theia  & 823K & 20.8M  & \multicolumn{1}{c|}{13.7M}  & 2.5M & 4.5M  \\ \hline
    E3-Trace  & 6M   & 274.8M & \multicolumn{1}{c|}{261.8M} & 813K & 12.2M \\ \hline
    OPTC      & 2M   & 11.2M  & \multicolumn{1}{c|}{174K}   & 7M   & 4M    \\ \hline
    \end{tabular}
  \end{table}
\SetKw{Continue}{continue}
  \begin{algorithm}[t!]
  \small
  \caption{Adaptive BFS Threat Graph Sampling}
  \label{adaptive bfs}
  \KwIn{Provenance graph $PPG$; Hop $k$; Indices of $POI_s$}
  \KwOut{Threat graphs $SG_s$}
  \SetAlgoLined
  % \SetAlgoNoEnd % This line disables the end after if statements
  $SG_s$ $\gets$ [] \\
  %\tcp{Load abstract node types that require heuristic sampling.}
  %$HST \gets LoadHeuriticSampleAbstractTypes() $\\
  \ForEach{poi $\in$ $POI_s$} {
      $S_{visited} \gets Set(poi)$ \\
      $Q_{nodes} \gets [(poi,0)]$ \\
      \ForEach{v,depth $\in Q_{nodes}$}{
        \tcp{Sampling using customized rules.}
        %\uIf{$v \in HST$}{
        %  $nei\_nodes \gets PPG.HeuristicSample(v)$  \\
        %}\uElse{
        %  $nei\_nodes \gets PPG.BiTraversal(v)$  \\
        %}
        $nei\_nodes \gets PPG.HeuristicSampling(v)$  \\

        \ForEach{u $\in nei\_nodes$}{
          \uIf{u $\in S_{visited}$ }{
            \Continue
          }
          $S_{visited}.add(u) $ \\ 
          \uIf{u $\in$ $POI_s$}{
              $Q_{nodes}.push((u,0)) $ \\ 
              \Continue
          }\uIf{PPG.Edges(u,v) \& $FORK\_MASK$}{
            $Q_{nodes}.push((u,depth)) $ \\
            }\uElseIf{depth < k - 1}{
              $Q_{nodes}.push((u,depth+1)) $ \\ 
          }
        }
      }
      $SG_s \gets SG_s + PPG.Subgraph(S_{visited})$ \\
  }
  \tcp{Merge subgraphs with overlapping nodes.}
  $SG_s \gets MergeSubgraphs(SG_s)$ \\
  \tcp{Merge identical nodes in threat graphs.}
  $SG_s \gets MergeIdenticalNodes(SG_s)$ \\
  \Return{$SG_s$} \\
  \end{algorithm}

\section{Graph Augmentation Techniques}\label{graph augmentation}
We use the following graph augmentation techniques to construct positive samples and enhance their diversity, with the perturbation ratio set to 20\%:
\begin{itemize}
  \item \textbf{Edge Perturbation.} This method involves randomly adding or removing edges specifically those of `Process $\to$ File' or `Process $\to$ NetFlow', as outlined in Table \ref{nodes edges}.
  \item \textbf{Node Perturbation.} This method introduces perturbations by randomly removing `File' or `NetFlow' nodes, or by incorporating nodes and edges from other graphs. 
\end{itemize}

\section{Threat Hunting Case}\label{case study}

We illustrate the workflow of \textsc{ProHunter} using the attack campaign from OpTC shown in Figure \ref{challenge}. The ``Malicious Upgrade'' campaign initiates with a \texttt{notepad++} update that downloads \texttt{update.exe}, a file containing a meterpreter payload. The attacker then gains system access via meterpreter, leverages \texttt{cmd.exe} to scan the local system, network, and shared resources. Finally, the attacker migrates to the system process \texttt{lsass.exe}, deploys \texttt{mimikatz} to gather credentials, and creates a root account for remote access. 

Assuming the PPG has been constructed, the threat hunting process proceeds as follows. First, upon parsing a batch of events, the PPG identified a POI event: `\texttt{lsass.exe} starts \texttt{cmd.exe}'. After integrating the current batch of events into the provenance graph, \textsc{ProHunter} executes the threat graph sampling algorithm with \texttt{lsass.exe} as the center node. During the sampling traversal, it visited several nodes related to historical POIs, including the \texttt{cKfGW.exe} file and the \texttt{firefox.exe} process. The algorithm then recursively executes sampling procedures centered on these historical POIs to capture anomalous interactions.
Upon completion of the traversal, \textsc{ProHunter} consolidates nodes with identical names and generates the comprehensive threat graph illustrated in Figure \ref{hunting example}.
\begin{figure}[]
  \centering
  \includegraphics[scale=1.2]{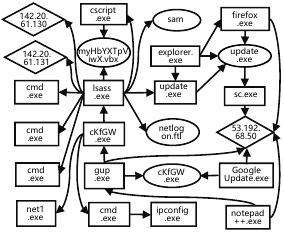}
  \caption{Sampled threat graph of OpTC: the attack ``Malicious Escalation'' starting on Day 3.}
  \label{hunting example}
\end{figure}
The resulting threat graph is subsequently transmitted to the attack representation and matching component for threat hunting. The attack representation model then encodes threat graph and all query graphs in parallel to generate robust attack representations. Finally, it calculates the semantic similarity and identifies the query graph with the highest similarity exceeding the threshold as the hunting result.

\section{Effects of Hyperparameter on Threat Hunting}\label{app:tuning exp}

\noindent\textbf{Sampling Algorithm.}
%MEGR-APT \cite{megrapt} sets the sampling hops based on the maximum path length among query graphs, which may result in some system entities being inaccessible. For example, the fork event in Linux systems extends the path of entity interactions, which may not be reflected in CTI reports. 
This experiment evaluates the performance of \textsc{ProHunter}'s sampling algorithm across different operating systems, with sampling hops ranging from 1 to 4.

The results shown in Figure~\ref{sampling tuning} demonstrate that as the sampling hop count increases, the coverage of anomalous nodes initially expands before reaching a plateau. However, this enhanced coverage comes at the cost of introducing more noisy nodes. For instance, at 4 sampling hops, the noise rates for both E3-Trace and E3-Cadets exceed 80\%.
%Additionally, the noise rate of E3-Trace is larger than any other datasets. Our in-depth analysis reveals that processes such as \texttt{firefox} and \texttt{cache} in E3-Trace make extensive use of the \textit{clone} system call, resulting in longer paths.
To achieve a balance between the coverage of anomalous interactions and the reduction of noise, we have chosen a sampling hop of 2 for all datasets.
%This phenomenon may cause the sampling algorithm of MEGR-APT \cite{megrapt} to fail to access some malicious nodes as it sets the sampling hop based on the maximum path length of CTI-derived query graphs, while the CTI report may miss these \textit{clone} or \textit{fork} operations, resulting in the final path length of the query graph being usually shorter than the actual path in the provenance graph.
\begin{figure}[]
  \centering
  \subfloat[\small{Coverage with varying $k$}]{\includegraphics[width=1.6in]{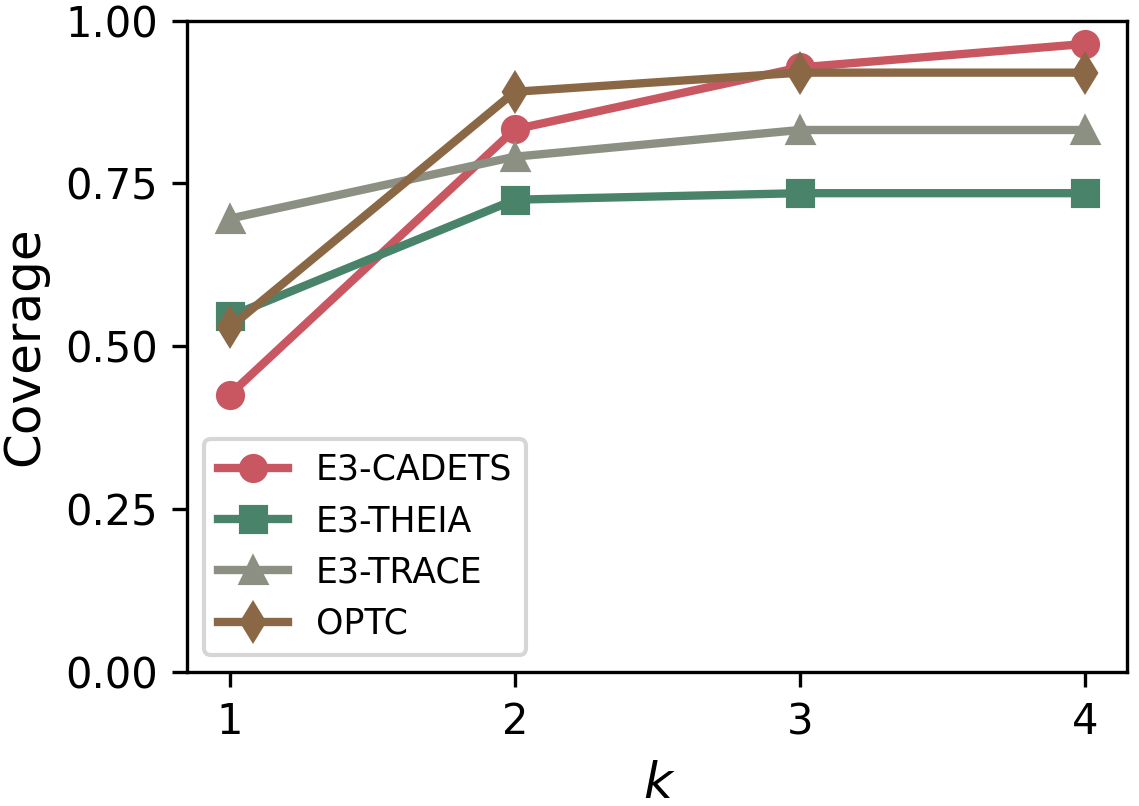}%
  \label{coverage}}
  \hfil
  \subfloat[\small{Noise rate with varying $k$}]{\includegraphics[width=1.6in]{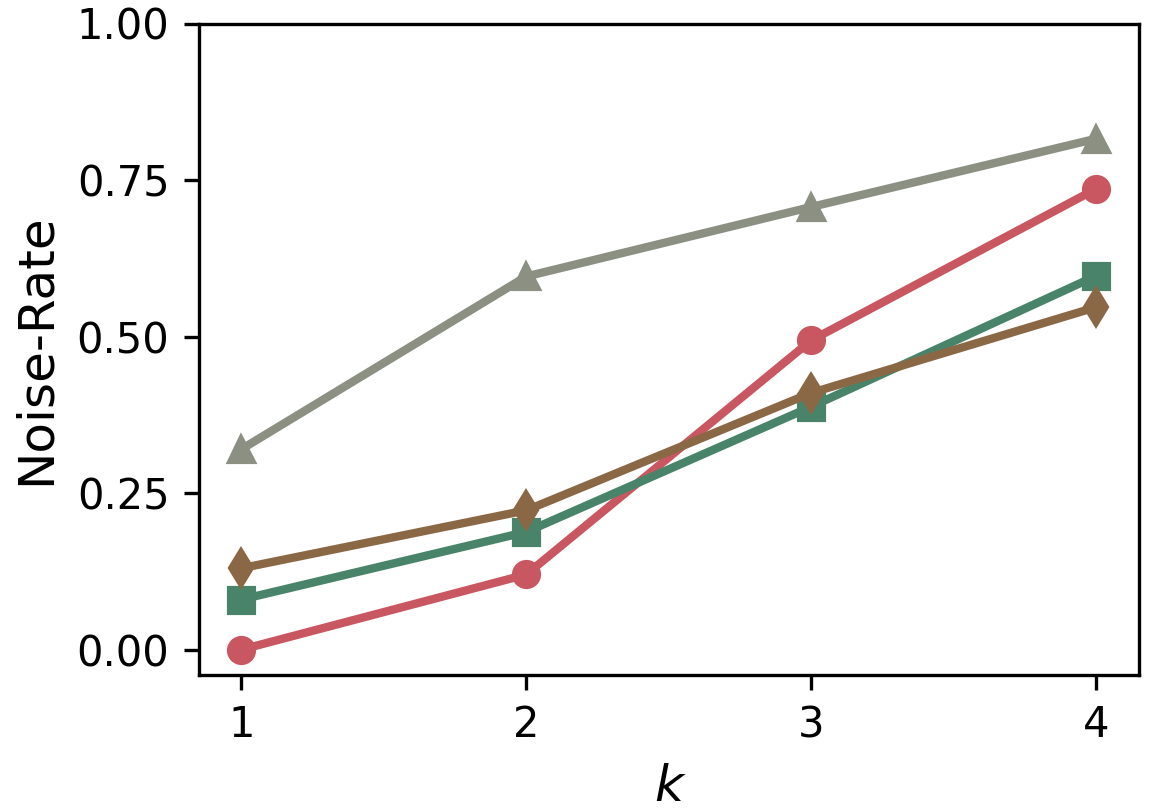}%
  \label{noise rate}}
  \caption{\centering{Effect of different sampling hops on threat graph.}}
  \label{sampling tuning}
\end{figure}
%This phenomenon may render some malicious entities inaccessible to the sampling algorithm of MEGR-APT \cite{megrapt}, which sets sampling hops based on the maximum path length of query graphs. Since the CTI reports may overlook these \textit{clone} or \textit{fork} operations, the resulting path length of query graphs is often shorter than the actual paths in provenance graphs.
%When setting the sampling hop, we prioritize high coverage followed by a low noise rate, as the absence of critical malicious interactions can result in incomplete attack semantics and threat hunting failures.
%Additionally, the adaptive graph representation of \textsc{ProHunter} enhances immunity to noisy nodes through the association of shared behaviors between matching graphs, as demonstrated in Section C.3. 
%The sampling hops selected for each dataset are shown in Table \ref{sampling analysis}.
%These results mean that the sampling hop $k$ should be tailored to different operating systems, we therefore set $k$=2 for Windows and FreeBSD, and $k$=4 for Linux.

\noindent\textbf{Graph Representation.}
\begin{figure}[t]
  \centering
  \subfloat[\small{AUC score with varying $l$}]{\includegraphics[width=1.5in]{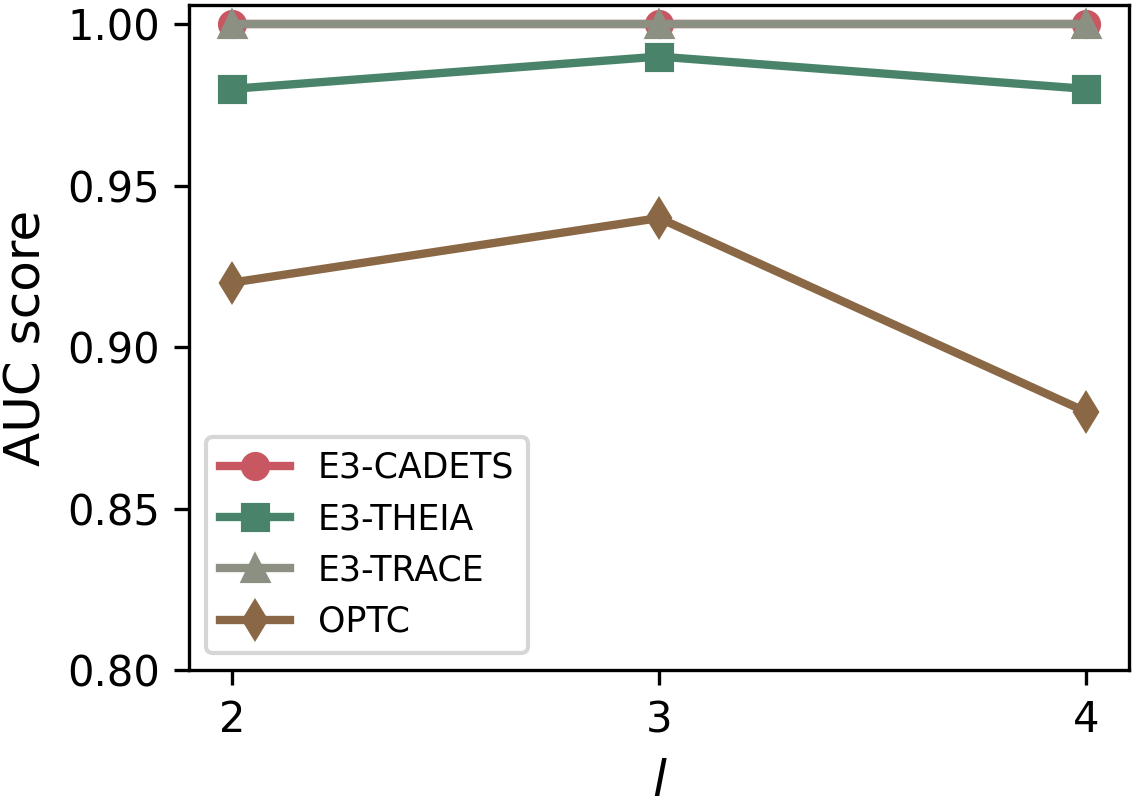}%
  \label{l}}
  \hfil
  \subfloat[\small{AUC score with varying $d$}]{\includegraphics[width=1.5in]{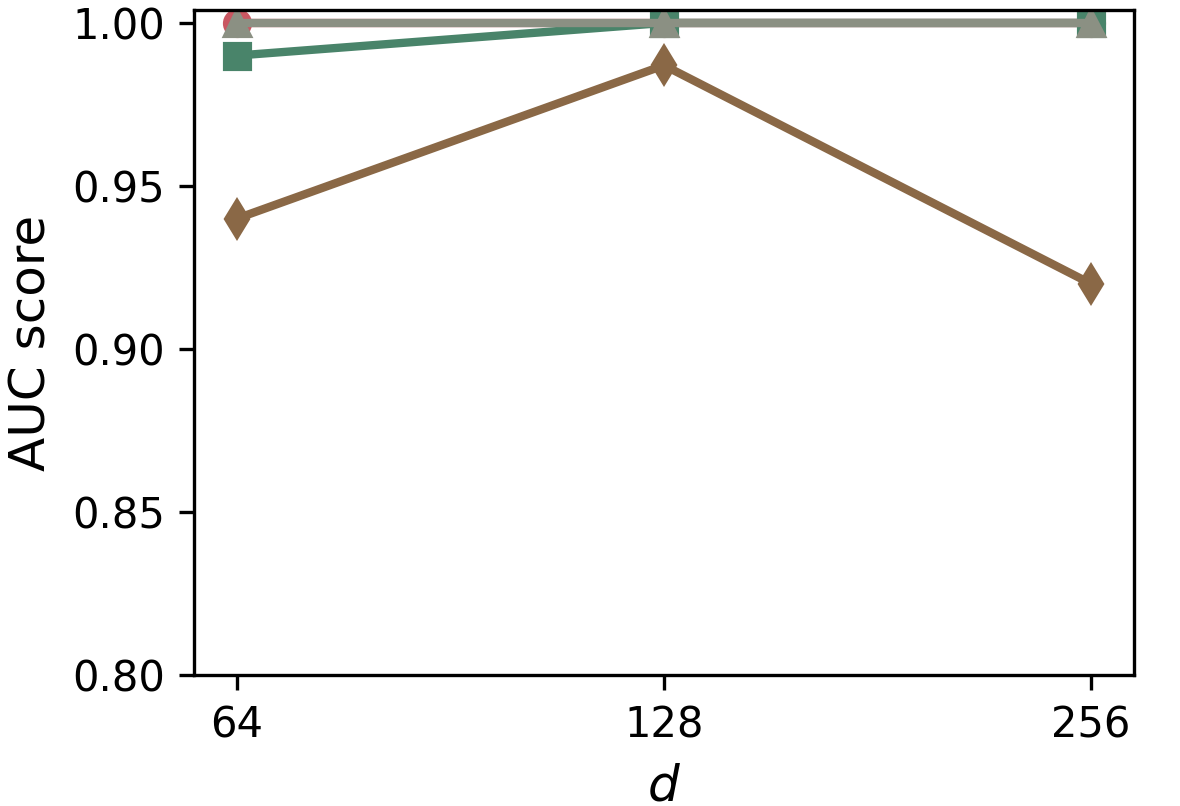}%
  \label{d}}
  \caption{Effect of different GNN's embedding sizes and message passing layers on ProHunter's performance.}
  \label{graph matching tuning}
\end{figure}
This experiment evaluates the impact of two key hyperparameters in the attack representation model on threat hunting performance: embedding size $d$ and the number of message passing layers $l$. First, we test $l$ between 2 and 4 with $d$ fixed at 64. Then, we fix $l$ at its optimal value and examine $d$ values ranging from 64 to 256.
Performance is quantified using the AUC score for threat hunting tasks, with experimental results illustrated in Figure~\ref{graph matching tuning}.

Analysis reveals that the E3 datasets are robust to variations in parameters $l$ and $d$, consistently achieving AUC scores near 1 across different parameter combinations. In contrast, performance on the OpTC dataset fluctuates, with optimal results achieved at $l$=3 and $d$=128. 
This suggests that larger vector dimensions and an increased number of message-passing layers enable the model to capture more complex behavior semantics. However, excessively large dimensions may introduce noise, ultimately degrading embedding quality. Balancing overhead and performance, we set $l$=3 and $d$=128 for all datasets.

\section{Point of Interest Nodes}\label{poi}
We adopt a Recall-first pattern matching method to identify anomalous nodes, which serve as POIs. Example patterns include `$<$firefox, write, *$>$' (i.e., browser process writes to files), `$<$*, read, /etc/group$>$' (i.e., a process accesses the credential file), and `$<$*, modify, /etc/hosts$>$' (i.e., a process modifies the system file). By real-time parsing of batch events and matching them against these sensitive operation patterns, we achieve both timely detection and high Recall. While this approach may result in numerous false positives, \textsc{ProHunter} can effectively filter these through attack semantic matching.
It is important to note that alerts generated by other advanced anomaly detection systems \cite{magic,flash,aptkgl} can also serve as POIs, but this is not the primary focus of our work.
\end{document}